\documentclass[a4paper, 11pt, oneside]{Thesis}  % Use the "Thesis" style, based on the ECS Thesis style by Steve Gunn
\graphicspath{Figures/}  % Location of the graphics files (set up for graphics to be in PDF format)

\usepackage[
  style=numeric-comp,
  backend=bibtex,
]{biblatex}
\usepackage{verbatim}  % Needed for the "comment" environment to make LaTeX comments
\usepackage{vector}  % Allows "\bvec{}" and "\buvec{}" for "blackboard" style bold vectors in maths
\usepackage{tabu}
\usepackage{bookmark}

\usepackage{listings}

\usepackage{wasysym}
\usepackage{pifont}% http://ctan.org/pkg/pifont
\newcommand{\cmark}[1][]{\(\CIRCLE_{#1}\)}% checkmark
\newcommand{\xmark}{\Circle}% crossmark
\newcommand{\omark}[1][]{\(\LEFTcircle_{#1}\)}% halfcheckmark

\usepackage{lscape}
\usepackage{rotating}
\usepackage[para]{threeparttable}
\usepackage{xpatch}
\makeatletter
\chardef\TPT@@@asteriskcatcode=\catcode`*
\catcode`*=11
\xpatchcmd{\threeparttable}
  {\TPT@hookin{tabular}}
  {\TPT@hookin{tabular}\TPT@hookin{tabu}}
  {}{} 
\catcode`*=\TPT@@@asteriskcatcode
\makeatother

\usepackage[utf8]{inputenc}

\usepackage[T1]{fontenc}
\usepackage[table]{xcolor}    % loads also »colortbl«

\hypersetup{urlcolor=blue, colorlinks=true}  % Colours hyperlinks in blue, but this can be distracting if there are many links.

\begin{document}
\frontmatter      % Begin Roman style (i, ii, iii, iv...) page numbering

% Set up the Title Page
\title  {Software solutions for form-based collection of data and the semantic enrichment of form data}
\authors  {\texorpdfstring
            {{Markus Steinberg}}
            {Markus Steinberg}
            }
\addresses  {\groupname\\\deptname\\\univname}  % Do not change this here, instead these must be set in the "Thesis.cls" file, please look through it instead
\date       {\today}
\subject    {}
\keywords   {}

\maketitle
%% ----------------------------------------------------------------

\setstretch{1.3}  % It is better to have smaller font and larger line spacing than the other way round

% Define the page headers using the FancyHdr package and set up for one-sided printing
\fancyhead{}  % Clears all page headers and footers
\rhead{\thepage}  % Sets the right side header to show the page number
\lhead{}  % Clears the left side page header

\pagestyle{fancy}  % Finally, use the "fancy" page style to implement the FancyHdr headers

\pagestyle{fancy}  %The page style headers have been "empty" all this time, now use the "fancy" headers as defined before to bring them back

%% ----------------------------------------------------------------
\lhead{\emph{Contents}}  % Set the left side page header to "Contents"
\tableofcontents  % Write out the Table of Contents

%% ----------------------------------------------------------------
%%\lhead{\emph{List of Figures}}  % Set the left side page header to "List if Figures"
%%\listoffigures  % Write out the List of Figures

%% ----------------------------------------------------------------
%%\lhead{\emph{List of Tables}}  % Set the left side page header to "List of Tables"
%%\listoftables  % Write out the List of Tables

%% ----------------------------------------------------------------
%%\setstretch{1.5}  % Set the line spacing to 1.5, this makes the following tables easier to read
%%\clearpage  % Start a new page
%%\lhead{\emph{Abbreviations}}  % Set the left side page header to "Abbreviations"
%%\listofsymbols{ll}  % Include a list of Abbreviations (a table of two columns)
{
% \textbf{Acronym} & \textbf{W}hat (it) \textbf{S}tands \textbf{F}or \\
%%\textbf{LAH} & \textbf{L}ist \textbf{A}bbreviations \textbf{H}ere \\

}

%% ----------------------------------------------------------------
%%\clearpage  % Start a new page
%%\lhead{\emph{Physical Constants}}  % Set the left side page header to "Physical Constants"
%%\listofconstants{lrcl}  % Include a list of Physical Constants (a four column table)
{
% Constant Name & Symbol & = & Constant Value (with units) \\
%%Speed of Light & $c$ & $=$ & $2.997\ 924\ 58\times10^{8}\ \mbox{ms}^{-\mbox{s}}$ (exact)\\

}

%% ----------------------------------------------------------------
%%\clearpage  %Start a new page
%%\lhead{\emph{Symbols}}  % Set the left side page header to "Symbols"
%%\listofnomenclature{lll}  % Include a list of Symbols (a three column table)
{
% symbol & name & unit \\
%%$a$ & distance & m \\
%%$P$ & power & W (Js$^{-1}$) \\
%%& & \\ % Gap to separate the Roman symbols from the Greek
%%$\omega$ & angular frequency & rads$^{-1}$ \\
}
%% ----------------------------------------------------------------
% End of the pre-able, contents and lists of things
% Begin the Dedication page

%%\setstretch{1.3}  % Return the line spacing back to 1.3

%%\pagestyle{empty}  % Page style needs to be empty for this page
%%\dedicatory{For/Dedicated to/To my\ldots}

%%\addtocontents{toc}{\vspace{2em}}  % Add a gap in the Contents, for aesthetics

%% ----------------------------------------------------------------
\mainmatter	  % Begin normal, numeric (1,2,3...) page numbering
\pagestyle{fancy}  % Return the page headers back to the "fancy" style
\lhead{}
% Include the chapters of the thesis, as separate files
% Just uncomment the lines as you write the chapters

\chapter*{Introduction}
\addcontentsline{toc}{chapter}{Introduction} 
Data collection is an important part of many citizen science projects as well as other fields of research, particularly in life sciences.
Mobile applications with form-based surveys are increasingly used to support this, due to the large number of mobile devices and their growing number of built-in sensors.
Since the composition of form-based surveys from scratch can be a tedious task, multiple tools have been published that can help with their design and distribution as well as the data collection via mobile devices and the data storage.
Some even support simple data analysis.
With this increasing number of software options project leaders will often face the question, which tool is most suitable for their current use case.

With that in mind, this student project pursues two main objectives:
\begin{enumerate}
\item To present an overview of a selection of survey design tools and their capabilities in order to provide a clear foundation for such a decision.
\item To examine if any tool provides the capability to collect and export data in a way that can easily be used and interpreted by other applications or persons.
This aspect includes the supply of metadata about the data collection process and the data itself, information about the meaning of the data as well as an export format that can easily be processed.
This objective is directly aiming towards a planned follow-up project.
\end{enumerate}

Chapter \ref{ch:formDescription} briefly covers the technical background by describing two form description standards that are relevant for the examined software tools.
In order to accomplish the stated goals, all of the tools were tested with respect to different aspects relevant to their usage.
First, a list of such aspects was compiled by comparing the features that are advertised by the tools themselves.
The coverage and usage of these features was then determined by thoroughly testing each of the tools:
First, general information about the tool like its open source repository or its license were identified by consulting its website and documentation.
Then, if no web-based version was offered, the tool was installed.
Afterwards a typical project's workflow was executed:
A form was designed, examining all available form-elements and form-building features like skip-logic or localization.
The survey was deployed to the mobile app, sample data was collected, submitted to the server and then exported.
In the end, additional features like visualization or data encryption were examined.
In cases where the availability of features was not obvious, the tool's community or its support were consulted for clarification.
The results for most of the examined aspects are illustrated in the form of feature-tables in chapter \ref{ch:comparison}.
Since some aspects are hard to cover in such feature-tables and require some ellaboration, chapter \ref{ch:descriptions} provides descriptions of the examined tools with regards to facets like general user-friendliness or required technical knowledge.

Of course, such a project will never be able to cover all published tools, especially since the number is evergrowing.
The tools that are examined here were selected because of 
    their rapid recent development (EpiCollect5\footnote{\url{https://five.epicollect.net}}),
    their large community (Open Data Kit 1 \& 2\footnote{\url{https://opendatakit.org/software/}}),
    explicit recommendations by the tool's users (Ohmage\footnote{\url{http://ohmage.org}}),
    their extensive feature repertoir (Kobo\footnote{\url{https://www.kobotoolbox.org}})
    or their professional design (SurveyCTO\footnote{\url{https://www.surveycto.com}} and Magpi\footnote{\url{https://home.magpi.com}}).
Other tools that are not covered here can of course be compared by examining the same aspects that are presented in this project.

\chapter{Form description standards}
\label{ch:formDescription}

This chapter offers brief descriptions of the two most important form-description standards:
The W3C\footnote{\url{https://www.w3.org/}} standard XForms (Section \ref{sec:xform}) and the XLSForm specification (Section \ref{sec:xlsform}).
XLSForms are used for form-authoring by many major data collection platforms while XForms is usually used as an internal format for the created forms.
Other standards like the old XHTML standard defined in \cite{w3c_xhtml} are not covered here, since they are not relevant to any of the examined data collection tools.

\section{XForm}
\label{sec:xform}
XForms is a standard that was published by the W3C.
Its current version, XForms 1.1, was published in \cite{w3c_xforms} in 2009.
The standard describes a markup language to be used for platform-independent form descriptions as well as an abstract processing model that should be implemented by applications that handle such descriptions.
While the technical details of the standard are out of scope for this project, the following aspects are especially relevant for mobile data collection:
XForms provides an explicit distinction between presentation, purpose and content of web forms, leading to a clear separation of the definition of control-elements and the data that is being collected \cite[Abstract]{w3c_xforms}.
Furthermore, the way in which a form control is rendered on a device or in a browser is not fixed \cite[2.1 An Example]{w3c_xforms}.
This implies, that a mobile device might render a certain form element different from a browser, specifically in a way that is more suitable for a small screen.

\begin{figure}
    \centering
    \includegraphics[width=0.7\textwidth]{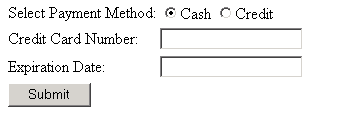}
        \caption{Example Web Form}
        \label{img:xformExample}
\end{figure}

Even with XForms defining a clearly separated structure the resulting form descriptions are rather complicated and technical.
An example, extracted from \cite[2.1 An Example]{w3c_xforms}, is shown in Listing \ref{lst:xformsModel}:
It illustrates the XForms model necessary to describe the simple web form shown in Figure \ref{img:xformExample}.
The input-controls for this form have to be defined separately in a markup language.
Listing \ref{lst:xhtmlControls} shows such controls described in XHTML.

\begin{lstlisting}[label=lst:xformsModel,caption=XForms model]
<xforms:model>
  <xforms:instance>
    <ecommerce xmlns="">
      <method/>
      <number/>
      <expiry/>
    </ecommerce>
  </xforms:instance>
  <xforms:submission action="http://example.com/submit" method="post" 
    id="submit" includenamespaceprefixes=""/>
</xforms:model>
\end{lstlisting}

\begin{lstlisting}[label=lst:xhtmlControls,caption=XHTML controls\, bound to the model]
<select1 ref="method">
  <label>Select Payment Method:</label>
  <item>
    <label>Cash</label>
    <value>cash</value>
  </item>
  <item>
    <label>Credit</label>
    <value>cc</value>
  </item>
</select1>
<input ref="number">
  <label>Credit Card Number:</label>
</input>
<input ref="expiry">
  <label>Expiration Date:</label>
</input>
<submit submission="submit">
  <label>Submit</label>
</submit>
\end{lstlisting}

While designing an XForms compatible XML-description of a form is possible, it would be a tedious undertaking since the descriptions quickly get longer and more complex the more features like validation or skip logic are used.
That is the reason why XLSForm, described in the next section, was developed.

\newpage
\section{XLSForm}
XLSForm is a form description standard "created to simplify the authoring of forms in Excel"\cite{xlsform}.
This goal is accomplished by providing a tool that can transform a Microsoft Excel file with a specific structure into an XForms compatible form description.
These Excel files have an intuitive structure so a form description can easily be built without any technical knowledge about XML or the XForms standard in particular.
The desired form elements are described in one worksheet while options for elements like single-choice-questions are defined in a separate worksheet.
This has two main advantages:
First, it provides a clear structure so the form author knows exactly where to look for which piece of information.
Secondly, it allows to refer to a certain set of possible answers multiple times; for example the option-set for "yes-or-no"-questions only has to be defined once.
An additional worksheet can be used to provide some general information about the form itself, like a title or a version number.

\begin{figure}
    \centering
    \includegraphics[width=0.8\textwidth]{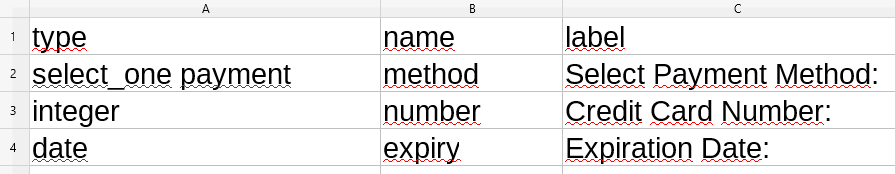}
        \caption{XLSForm Example - Survey worksheet}
        \label{img:xlsformSurvey}
\end{figure}

\begin{figure}
    \centering
    \includegraphics[width=0.6\textwidth]{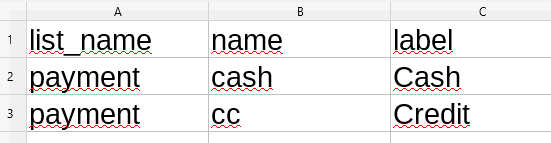}
        \caption{XLSForm Example - Choices worksheet}
        \label{img:xlsformChoices}
\end{figure}

\begin{figure}
    \centering
    \includegraphics[width=0.75\textwidth]{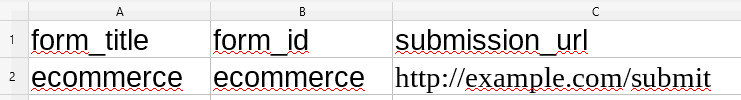}
        \caption{XLSForm Example - Settings worksheet}
        \label{img:xlsformSettings}
\end{figure}

Figures \ref{img:xlsformSurvey}, \ref{img:xlsformChoices} and \ref{img:xlsformSettings} show the three worksheets that describe the web form in Figure \ref{img:xformExample}.
These worksheets are evidently more human readable and easier to modify than the corresponding XForms description.

XLSForms does not provide support for the full XForms standard, it is only compatible with a subset of its features.
This subset was defined by the Javarosa project\footnote{\url{https://bitbucket.org/javarosa/javarosa/wiki/Home}}.
Since the project has been inactive, the original server that was hosting most of its information, including the subset definition, has been shut down.
However, the XLSForms standard, and with it the XForms subset, has been adopted by the Open Data Kit community in their ODK XForms Specification\footnote{\url{http://opendatakit.github.io/xforms-spec/}}.
XLSForms has since become a widely accepted form description standard and is used by multiple data collection platforms.
\label{sec:xlsform}
\chapter{Comparison}
\label{ch:comparison}

This chapter provides an assessment of different aspects of interest offered by the examined tools.
First, brief explanations are given for each aspect.
Afterwards, tables illustrate which tools include which features.
To give a better overview, these tables are separated into multiple categories:
Table \ref{tab:form_elements} contains information about form-elements that are available to build a survey.
Multiple tables then display different examined features:
Steps of a data collection workflow that are supported by the tools (Table \ref{tab:features_workflowSteps}), Features enhancing the survey design (Table \ref{tab:features_surveyDesign}), the form description standard compliance (Table \ref{tab:features_formDescription}), data export formats (Table \ref{tab:features_dataExport}), development related information (Table \ref{tab:features_development}) and other features of interest (Table \ref{tab:features_misc}).
At the end of this chapter, one last table presents the metadata that can be collected with each of the tools (Table \ref{tab:metadata}).

\section{Form-Elements}
This section provides an overview of the different form-elements that can be used to build a survey.

\begin{itemize}
\item Text: Simple text input.
\item Integer: Whole number input.
\item Double: Decimal number input.
\item Date: Selection of a date, typically using a calendar or spinners.
\item Time: Selection of a time of day, typically using spinners.
\item Single-choice-question: Selection of a single option from a defined set of options. Typically using radio buttons or a dropdown-list.
\item Multiple-choice-question: Selection of one or more options from a defined set of options. Typically using checkboxes.
\item Readme: Display of arbitrary text to the user. No input.
\item Location (Manually): Manual input of geographical information. Typically by placing a pointer on a map or explicitly stating latitude, longitude, altitude and accuracy.
\item Location (Auto): Automated detection of the current location, typically using GPS or GSM.
\item Path: Input of multiple geolocations connected via a path.
\item Area: Input of a geographical area, typically defined by a polygon or a bounding box.
\item Image: Upload an image from device storage or take a new picture using the collection devices camera.
\item Audio: Upload or record an audio signal.
\item Video: Upload or record a video.
\item Barcode: Scan a barcode and collect the encoded information.
\item Rating: Assign defined categories (e.g. "good", "neutral" and "bad") to a defined set of options.
\item Ranking: Rank a defined set of options by assigning an order.
\item NFC: Read information via Near-Field Communication\footnote{\url{http://nearfieldcommunication.org/about-nfc.html}}.
\end{itemize}

%Form-elements table
\begin{sidewaystable}[h!tbp]
	\setlength\extrarowheight{1pt} % provide a bit more vertical whitespace
	\centering
	\footnotesize
	\captionsetup{size=footnotesize}
	\begin{threeparttable}
		\caption{The form elements included in the examined tools. \\
		\hspace{1.3cm} (\cmark \ldots element included; \omark \ldots element partially included; \xmark \ldots element not included)}
		\label{tab:form_elements}
		\tabulinesep=0.4mm % http://tex.stackexchange.com/a/64338/63780
		\rowcolors{2}{gray!25}{white}
\begin{tabu} to \linewidth{X[2,r,m]|X[1,m,c]X[1,m,c]X[1,m,c]X[1,m,c]X[1,m,c]X[1,m,c]X[1,m,c]} 
\toprule
\rowcolor{gray!50}
\strut  & \strut EpiCollect 5 & \strut ODKv1 & \strut ODKv2 & \strut Kobo & \strut Ohmage & \strut SurveyCTO & \strut Magpi \\ 
Text & \strut \cmark & \strut \cmark & \strut \cmark & \strut \cmark & \strut \cmark & \strut \cmark & \strut \cmark \\ 
Numeric (Integer) & \strut \cmark & \strut \cmark & \strut \cmark & \strut \cmark & \strut \cmark & \strut \cmark & \strut \cmark \\ 
Decimal & \strut \cmark & \strut \cmark & \strut \cmark & \strut \cmark & \strut \cmark & \strut \cmark & \strut \cmark \\ 
Date & \strut \cmark & \strut \cmark & \strut \cmark & \strut \cmark & \strut \cmark & \strut \cmark & \strut \cmark \\ 
Time & \strut \cmark & \strut \cmark & \strut \cmark & \strut \cmark & \strut \cmark & \strut \cmark & \strut \cmark \\ 
Single-choice-question & \strut \cmark & \strut \cmark & \strut \cmark & \strut \cmark & \strut \cmark & \strut \cmark & \strut \cmark \\ 
Multiple-choice-question & \strut \cmark & \strut \cmark & \strut \cmark & \strut \cmark & \strut \cmark & \strut \cmark & \strut \cmark \\ 
Readme & \strut \cmark & \strut \cmark & \strut \cmark & \strut \cmark & \strut \cmark & \strut \cmark & \strut \cmark \\ 
Location (Manually) & \strut \xmark & \strut \cmark & \strut \xmark & \strut \xmark & \strut \xmark & \strut \cmark & \strut \xmark \\ 
Location (Auto) & \strut \cmark & \strut \cmark & \strut \cmark & \strut \cmark & \strut \omark \tnote{a} & \strut \cmark & \strut \cmark \\ 
Path & \strut \xmark & \strut \xmark & \strut \xmark & \strut \cmark & \strut \xmark & \strut \cmark & \strut \xmark \\ 
Area & \strut \xmark & \strut \xmark & \strut \xmark & \strut \cmark & \strut \xmark & \strut \cmark & \strut \xmark \\ 
Image & \strut \cmark & \strut \cmark & \strut \cmark & \strut \cmark & \strut \cmark & \strut \cmark & \strut \xmark \\ 
Audio & \strut \cmark & \strut \cmark & \strut \cmark & \strut \cmark & \strut \xmark & \strut \cmark & \strut \xmark \\ 
Video & \strut \cmark & \strut \cmark & \strut \cmark & \strut \cmark & \strut \xmark & \strut \cmark & \strut \xmark \\ 
Barcode & \strut \cmark & \strut \cmark & \strut \cmark & \strut \cmark & \strut \xmark & \strut \cmark & \strut \xmark \\ 
Rating & \strut \xmark & \strut \xmark & \strut \xmark & \strut \cmark & \strut \xmark & \strut \xmark & \strut \xmark \\ 
Ranking & \strut \xmark & \strut \xmark & \strut \xmark & \strut \cmark & \strut \xmark & \strut \xmark & \strut \xmark \\ 
NFC & \strut \xmark & \strut \xmark & \strut \xmark & \strut \xmark & \strut \xmark & \strut \xmark & \strut \cmark \\ 
\bottomrule
 \end{tabu}
 
\begin{tablenotes}
\item[a]{Location is automatically submitted as part of the metadata}
\end{tablenotes}
	\end{threeparttable}
\end{sidewaystable}

\newpage
\section{Features}

\paragraph*{Workflow steps}\mbox{}\\
This section provides an overview of the project workflow steps that are supported by the different data collection tools.
\begin{itemize}
\item Form design: Is the design of a data collection survey supported by the tool?
\item Data collection: Is data collection supported by the tool?
\item Visualization: Which visualization types are provided by the tool?
\item Analysis: Does the tool provide support for basic data analysis?
\item Publication: Which cloud platforms and tools can data be published to directly?
\end{itemize}

\paragraph*{Survey design}\mbox{}\\
This section illustrates different features that simplify form-authoring or allow to build more sophisticated and user-friendly surveys.
\begin{itemize}
\item Form Designer: Build a survey in a user interface, typically using drag and drop.
\item Skip Logic: Skip certain parts of a survey, depending on previous answers.
\item Localization: Define labels for questions in multiple languages so the survey can automatically be translated to a user's preferred language.
\item Calculations: Evaluate mathematical or logical expressions referencing previous answers during the survey and use the results in skip logic, readmes, ...
\item Queries\footnote{Detailed usage description: \url{https://docs.opendatakit.org/odk2/xlsx-converter-using/\#using-queries} }: Read data from a structured source (e.g. CSV files) and use the results for skip logic, as answer-options, ...
\item Linked Tables\footnote{Detailed usage description: \url{https://docs.opendatakit.org/odk2/xlsx-converter-using/\#linked-tables}}: Launch subforms that store data in different tables.
\item Required \& Optional fields: Mark a question as required or optional to indicate if the survey can be finished without providing an answer.
\item Validation: Define validity constrains for form-fields, e.g. a range of valid values for a number input.
\end{itemize}

\paragraph*{Form description standards}\mbox{}\\
This section presents which of the form description standards mentioned in chapter \ref{ch:formDescription} are supported when trying to import or export designed forms.
\begin{itemize}
\item XForms Export: Export forms compatible with the XForms standard.
\item XForms Import: Import forms compatible with the XForms standard.
\item XLSForm Export: Export forms in the XLSForm format.
\item XLSForm Import: Import forms in the XLSForm format.
\end{itemize}

\paragraph*{Data export formats}\mbox{}\\
This section lists the different file formats in which collected data can be exported.
This refers only to file export, not to the publication on cloud platform mentioned in Table \ref{tab:features_workflowSteps}.
\begin{itemize}
\item CSV Export: Export collected data in CSV format.
\item JSON Export: Export collected data in JSON format.
\item XLS Export: Export collected data in XLS format.
\item XML Export: Export collected data in XML format.
\item XML/KML Export: Export collected geographical data in KML format.
\item RDF Export: Export collected data in RDF format.
\end{itemize}

\paragraph*{Development related\mbox{}\\}
This section contains aspects that are most relevant to developers who want to extend the existing tools and to projects leaders who want to set up their own instance of one of the tools.
\begin{itemize}
\item Active Development: Is the software currently under active development? This is judged, if possible, by Git-commits in the past six months, otherwise by activity on social media or in forums.
\item License: License under which the software is provided.
\item Open Source: Is the full software provided open source? This includes all parts that are required to build and deploy a survey, collect data with a mobile device and store the data on a server.
\item Programming language: Programming languages that are used for the tool's development.
\item Self-hosting: Is it possible to host the software yourself so the collected data is stored on your own server?
\end{itemize}

\paragraph*{Miscellaneous features\mbox{}\\}
This section consists of different aspects that are potentially important to evaluate the fitness for use of the tools but don't fit into any of the previous categories.
\begin{itemize}
\item Data Encryption: Can submitted data be encrypted and stay encrypted while stored on the server? This requires the form author to provide a public key and to later decrypt the data using a private key. This feature does not only refer to the data transport, so SSL support for data submission is not sufficient.  
\item Semantic Enrichment: Does the software allow to provide some kind of semantic enrichment (e.g. mapping certain form-fields to ontology terms) or does it provide semantic enrichtment out-of-the-box?
\item HXL support: Map form-fields to tags and attributes of the Humanitarian Exchange Language (HXL), which is a data standard "designed to improve information sharing during a humanitarian crisis"\cite{hxlDoc}. 
\item Building custom prompts: Build prompts with custom functionality, typically using a markup language like HTML for the presentation and a programming language to define the functionality.
\item Visualization \& Data management on mobile device: Visualize (e.g. show on a map) and edit data entries on the mobile device.
\item Offline-Collection: Can data be collected without active internet access? Data can be uploaded to a server at a later point in time.
\item Mobile OS support: Which operating systems are supported by the mobile app and, for open source tools, whether the app was developed natively or platform independent.
\end{itemize}

%Workflow steps table
%Survey-design features table
\begin{sidewaystable}
	\setlength\extrarowheight{1pt} % provide a bit more vertical whitespace
	\centering
	\footnotesize
	\captionsetup{size=footnotesize}
	\begin{threeparttable}
		\caption{The workflow steps supported by the examined tools. \\
		\hspace{1.3cm} (\cmark \ldots step supported; \omark \ldots step partially supported; \xmark \ldots step not supported) }
		\label{tab:features_workflowSteps}
		\tabulinesep=0.4mm % http://tex.stackexchange.com/a/64338/63780
		\rowcolors{2}{gray!25}{white}
\begin{tabu} to \linewidth{X[3,r,m]|X[1,m,c]X[1,m,c]X[1,m,c]X[1,m,c]X[1,m,c]X[1,m,c]X[1,m,c]}
\toprule
\rowcolor{gray!50}
\strut  & \strut EpiCollect 5 & \strut ODKv1 & \strut ODKv2 & \strut Kobo & \strut Ohmage & \strut SurveyCTO & \strut Magpi \\
\hline
Form design & \strut \cmark & \strut \cmark & \strut \omark\tnote{a} & \strut \cmark & \strut \cmark & \strut \cmark & \strut \cmark \\
Data collection & \strut \cmark & \strut \cmark & \strut \cmark & \strut \cmark & \strut \cmark & \strut \cmark & \strut \cmark \\
Visualization & \strut Map, \mbox{Pie chart} & \strut Map, \mbox{Pie chart}, \mbox{Bar chart} & \strut Map & \strut Map, \mbox{Pie chart}, \mbox{Bar chart}, \mbox{Line chart}, \mbox{Area chart} & \strut Map, \mbox{Pie chart}, \mbox{Bar chart}, \mbox{Line chart} & \strut Map, \mbox{Pie chart}, \mbox{Bar chart}, \mbox{Scatterplot}, \mbox{Trend plot\tnote{b}} & \strut Map \\
Analysis & \strut \xmark & \strut \xmark & \strut \xmark & \strut \xmark & \strut \xmark & \strut \xmark & \strut \xmark \\
Publication & \strut Google Spreadsheet (via API) & \strut  Google Spreadsheet, Google FusionTables, REDCap Server\tnote{c}, Custom JSON Server & \strut \xmark & \strut \xmark & \strut \xmark & \strut Google Spreadsheet, Google FusionTables, Zapier\tnote{d} & \strut \xmark \\
\bottomrule
\end{tabu}

\begin{tablenotes}
\item[a]{Only conversion of XLSForm to ODK2 application format}
\item[b]{Plotting a numeric value over time}
\item[c]{https://www.project-redcap.org}
\item[d]{https://zapier.com}
\end{tablenotes}

	\end{threeparttable}
\end{sidewaystable}

%Survey-design features table
\begin{sidewaystable}
	\setlength\extrarowheight{1pt} % provide a bit more vertical whitespace
	\centering
	\footnotesize
	\captionsetup{size=footnotesize}
	\begin{threeparttable}
		\caption{The survey design features provided by the examined tools. \\
		\hspace{1.3cm} (\cmark \ldots feature included; \omark \ldots feature partially included; \xmark \ldots feature not included) }
		\label{tab:features_surveyDesign}
		\tabulinesep=0.4mm % http://tex.stackexchange.com/a/64338/63780
		\rowcolors{2}{gray!25}{white}
\begin{tabu} to \linewidth{X[3,r,m]|X[1,m,c]X[1,m,c]X[1,m,c]X[1,m,c]X[1,m,c]X[1,m,c]X[1,m,c]}
\toprule
\rowcolor{gray!50}
\strut  & \strut EpiCollect 5 & \strut ODKv1 & \strut ODKv2 & \strut Kobo & \strut Ohmage & \strut SurveyCTO & \strut Magpi \\
\hline
Form Designer & \strut \cmark & \strut \cmark & \strut \xmark & \strut \cmark & \strut \cmark & \strut \cmark & \strut \cmark \\
Skip Logic & \strut \cmark & \strut \cmark & \strut \cmark & \strut \cmark & \strut \xmark & \strut \cmark & \strut \cmark \\ 
Localization & \strut \xmark & \strut \cmark & \strut \cmark & \strut \omark\tnote{a} & \strut \xmark & \strut \cmark & \strut \xmark \\
Calculations & \strut \xmark & \strut \cmark & \strut \cmark & \strut \xmark & \strut \xmark & \strut \cmark & \strut \cmark \\
Queries & \strut \xmark & \strut \xmark & \strut \cmark & \strut \xmark & \strut \xmark & \strut \xmark & \strut \xmark \\
Linked Tables & \strut \xmark & \strut \xmark & \strut \cmark & \strut \xmark & \strut \xmark & \strut \xmark & \strut \xmark \\
Required \& Optional Fields & \strut \cmark & \strut \cmark & \strut \cmark & \strut \cmark & \strut \cmark & \strut \cmark & \strut \cmark \\
Validation & \strut \cmark & \strut \cmark & \strut \cmark & \strut \cmark & \strut \cmark & \strut \cmark & \strut \cmark \\
\bottomrule
\end{tabu}

\begin{tablenotes}
\item[a]{Not supported in form designer, has to be added manually in .xls file}
\end{tablenotes}

	\end{threeparttable}
\end{sidewaystable}
	
%Form description standards features table
\begin{sidewaystable}
	\setlength\extrarowheight{1pt} % provide a bit more vertical whitespace
	\centering
	\footnotesize
	\captionsetup{size=footnotesize}
	\begin{threeparttable}
		\caption{The form description formats supported by the examined tools. \\
		\hspace{1.3cm} (\cmark \ldots feature included; \omark \ldots feature partially included; \xmark \ldots feature not included) }
		\label{tab:features_formDescription}
		\tabulinesep=0.4mm % http://tex.stackexchange.com/a/64338/63780
		\rowcolors{2}{gray!25}{white}
\begin{tabu} to \linewidth{X[2,r,m]|X[1,m,c]X[1,m,c]X[1,m,c]X[1,m,c]X[1,m,c]X[1,m,c]X[1,m,c]} 
\toprule
\rowcolor{gray!50}
\strut  & \strut EpiCollect 5 & \strut ODKv1 & \strut ODKv2 & \strut Kobo & \strut Ohmage & \strut SurveyCTO & \strut Magpi \\ 
\hline 
XForms Export & \strut \xmark & \strut \omark\tnote{a} & \strut \xmark & \strut \omark \tnote{a} & \strut \xmark & \strut \omark \tnote{a} & \strut \xmark \\ 
XForms Import & \strut \xmark & \strut \xmark & \strut \xmark & \strut \xmark & \strut \xmark & \strut \omark \tnote{a} & \strut \cmark \\ 
XLSForm Export & \strut \xmark & \strut \cmark & \strut \xmark & \strut \cmark & \strut \xmark & \strut \cmark & \strut \xmark \\ 
XLSForm Import & \strut \xmark & \strut \xmark & \strut \cmark & \strut \cmark & \strut \xmark & \strut \cmark & \strut \xmark \\ 
\bottomrule
\end{tabu}

\begin{tablenotes}
\item[a]{Using the ODK XForms Specification defined in \cite{odkXFormSpec}, a subset of the W3C XForms 1.0 specification.}
\end{tablenotes}
	\end{threeparttable}
\end{sidewaystable}

%Data export format features table
\begin{sidewaystable}
	\setlength\extrarowheight{1pt} % provide a bit more vertical whitespace
	\centering
	\footnotesize
	\captionsetup{size=footnotesize}
	\begin{threeparttable}
		\caption{The data export formats supported by the examined tools. \\
		\hspace{1.3cm} (\cmark \ldots feature included; \xmark \ldots feature not included) }
		\label{tab:features_dataExport}
		\tabulinesep=0.4mm % http://tex.stackexchange.com/a/64338/63780
		\rowcolors{2}{gray!25}{white}
\begin{tabu} to \linewidth{X[2,r,m]|X[1,m,c]X[1,m,c]X[1,m,c]X[1,m,c]X[1,m,c]X[1,m,c]X[1,m,c]} 
\toprule
\rowcolor{gray!50}
\strut  & \strut EpiCollect 5 & \strut ODKv1 & \strut ODKv2 & \strut Kobo & \strut Ohmage & \strut SurveyCTO & \strut Magpi \\ 
\hline 
CSV Export & \strut \cmark & \strut \cmark & \strut \cmark & \strut \cmark & \strut \cmark & \strut \cmark & \strut \cmark \\ 
JSON Export & \strut \cmark & \strut \cmark & \strut \xmark & \strut \xmark & \strut \xmark & \strut \xmark & \strut \xmark \\ 
XLS Export & \strut \xmark & \strut \xmark & \strut \xmark & \strut \cmark & \strut \xmark & \strut \cmark & \strut \cmark \\ 
XML Export & \strut \xmark & \strut \xmark & \strut \xmark & \strut \xmark & \strut \xmark & \strut \xmark & \strut \xmark \\ 
XML/KML Export & \strut \xmark & \strut \cmark & \strut \xmark & \strut \xmark & \strut \xmark & \strut \cmark & \strut \xmark \\ 
RDF Export & \strut \xmark & \strut \xmark & \strut \xmark & \strut \xmark & \strut \xmark & \strut \xmark & \strut \xmark \\ 
\bottomrule
\end{tabu}
	\end{threeparttable}
\end{sidewaystable}

%Development related features table
\begin{sidewaystable}
	\setlength\extrarowheight{1pt} % provide a bit more vertical whitespace
	\centering
	\footnotesize
	\captionsetup{size=footnotesize}
	\begin{threeparttable}
		\caption{The development related features provided by the examined tools. \\
		\hspace{1.3cm} (\cmark \ldots feature included; \xmark \ldots feature not included) }
		\label{tab:features_development}
		\tabulinesep=0.4mm % http://tex.stackexchange.com/a/64338/63780
		\rowcolors{2}{gray!25}{white}
\begin{tabu} to \linewidth{X[2,r,m]|X[1,m,c]X[1,m,c]X[1,m,c]X[1,m,c]X[1,m,c]X[1,m,c]X[1,m,c]} 
\toprule
\rowcolor{gray!50}
\strut  & \strut EpiCollect 5 & \strut ODKv1 & \strut ODKv2 & \strut Kobo & \strut Ohmage & \strut SurveyCTO & \strut Magpi \\ 
\hline 
Active Development & \strut \cmark & \strut \cmark & \strut \cmark & \strut \cmark & \strut \xmark & \strut \cmark & \strut \cmark \\ 
Open Source & \strut \xmark & \strut \cmark & \strut \cmark & \strut \cmark & \strut \cmark & \strut \xmark & \strut \xmark \\ 
Programming language & \strut - & \strut Java, JavaScript, Python & \strut Java, JavaScript & \strut Java, JavaScript, Python & \strut Java, \mbox{Objective C} & \strut - & \strut - \\
License & \strut - & \strut Apache\tnote{a} & \strut Apache\tnote{a} & \strut Apache\tnote{a} \hspace{1pt}, GNU\tnote{b} & \strut Apache\tnote{a} & \strut - & \strut - \\ 
Self-hosting & \strut \xmark & \strut \cmark & \strut \cmark & \strut \cmark & \strut \cmark & \strut \xmark & \strut \xmark \\ 
\bottomrule
\end{tabu}

\begin{tablenotes}
\item[a]{Apache License 2.0}
\item[b]{GNU Affero General Public License v3.0}
\end{tablenotes}
	\end{threeparttable}
\end{sidewaystable}

%Miscellaneous features table
\begin{sidewaystable}
	\setlength\extrarowheight{1pt} % provide a bit more vertical whitespace
	\centering
	\footnotesize
	\captionsetup{size=footnotesize}
	\begin{threeparttable}
		\caption{Other features provided by the examined tools. \\
		\hspace{1.3cm} (\cmark \ldots feature included; \omark \ldots feature partially included; \xmark \ldots feature not included) }
		\label{tab:features_misc}
		\tabulinesep=0.4mm % http://tex.stackexchange.com/a/64338/63780
		\rowcolors{2}{gray!25}{white}
\begin{tabu} to \linewidth{X[3,r,m]|X[1,m,c]X[2,m,c]X[2,m,c]X[2,m,c]X[2,m,c]X[1,m,c]X[1,m,c]} 
\toprule
\rowcolor{gray!50}
\strut  & \strut EpiCollect 5 & \strut ODKv1 & \strut ODKv2 & \strut Kobo & \strut Ohmage & \strut SurveyCTO & \strut Magpi \\ 
\hline 
Data encryption & \strut \xmark & \strut \cmark & \strut \xmark & \strut \cmark & \strut \xmark & \strut \cmark & \strut \xmark \\ 
Semantic Enrichment & \strut \xmark & \strut \xmark & \strut \xmark & \strut \xmark & \strut \xmark & \strut \xmark & \strut \xmark \\ 
HXL support & \strut \xmark & \strut \xmark & \strut \xmark & \strut \cmark & \strut \xmark & \strut \xmark & \strut \xmark \\ 
Building custom prompts & \strut \xmark & \strut \xmark & \strut \cmark & \strut \xmark & \strut \xmark & \strut \xmark & \strut \xmark \\ 
Visualization/Data management on mobile device & \strut \xmark & \strut \xmark & \strut \cmark & \strut \xmark & \strut \xmark & \strut \xmark & \strut \omark \tnote{a} \\ 
Offline-Collection & \strut \cmark & \strut \cmark & \strut \cmark & \strut \cmark & \strut \cmark & \strut \cmark & \strut \cmark \\ 
Mobile OS support & \strut Android, iOS & \strut \mbox{Android (native)} & \strut \mbox{Android (native)} & \strut \mbox{Android (native)} & \strut \mbox{Android (native)}, \mbox{iOS (native)} & \strut Android & \strut Android, iOS \\
\bottomrule
\end{tabu}

\begin{tablenotes}
\item[a]{Geographical data only.}
\end{tablenotes}
	\end{threeparttable}
\end{sidewaystable}

\newpage
\section{Metadata}
\paragraph*{Metadata\mbox{}\\}
This section describes the different kinds of metadata that can be gathered by the data collection tools
\begin{itemize}
\item Start: Timestamp or date when an instance of the survey was started on the collection device.
\item End: Timestamp or date when an instance of the survey was finalized on the collection device.
\item Today: Timestamp or date when an instance of the survey was submitted to the data server.
\item Device-ID: International Mobile Equipment Identity (IMEI) - Unique identifier for mobile phones.
\item SIM serial: Identifier of the Subscriber Identity Module (SIM) used in the collection device.
\item Subscriber-ID: International Mobile Subscriber Identity (IMSI) - Unique identifier for mobile network users, usually stored on the SIM card.
\item Phone number: Phone number of the collection device.
\item Username: Name of the user account, registered on the data collection platform.
\item Email: Email-address used for the account registered on the data collection platform.
\item Duration: Duration from starting the survey to finalizing.
\end{itemize}

%Metadata table
\begin{sidewaystable}
	\setlength\extrarowheight{1pt} % provide a bit more vertical whitespace
	\centering
	\footnotesize
	\captionsetup{size=footnotesize}
	\begin{threeparttable}
		\caption{The metadata collection features provided by the examined tools. \\
		\hspace{1.3cm} (\cmark \ldots feature included; \xmark \ldots feature not included) }
		\label{tab:metadata}
		\tabulinesep=0.4mm % http://tex.stackexchange.com/a/64338/63780
		\rowcolors{2}{gray!25}{white}
\begin{tabu} to \linewidth{X[2,r,m]|X[1,m,c]X[1,m,c]X[1,m,c]X[1,m,c]X[1,m,c]X[1,m,c]X[1,m,c]} 
\toprule
\rowcolor{gray!50}
\strut  & \strut EpiCollect 5 & \strut ODKv1 & \strut ODKv2 & \strut Kobo & \strut Ohmage & \strut SurveyCTO & \strut Magpi \\ 
\hline 
Start (time \& date) & \strut \xmark & \strut \cmark & \strut \xmark & \strut \cmark & \strut \xmark & \strut \cmark & \strut \cmark \\ 
End (time \& date) & \strut \xmark & \strut \cmark & \strut \xmark & \strut \cmark & \strut \xmark & \strut \cmark & \strut \cmark \\ 
Today (day of the survey) & \strut \cmark & \strut \cmark & \strut \cmark & \strut \cmark & \strut \cmark & \strut \xmark & \strut \cmark \\ 
Device-ID (IMEI) & \strut \xmark & \strut \cmark & \strut \xmark & \strut \cmark & \strut \xmark & \strut \cmark & \strut \xmark \\ 
Subscriber-ID (IMSI) & \strut \xmark & \strut \cmark & \strut \xmark & \strut \cmark & \strut \xmark & \strut \cmark & \strut \xmark \\ 
SIM serial & \strut \xmark & \strut \cmark & \strut \xmark & \strut \cmark & \strut \xmark & \strut \cmark & \strut \xmark \\ 
Phone number & \strut \xmark & \strut \cmark & \strut \xmark & \strut \cmark & \strut \xmark & \strut \cmark & \strut \xmark \\ 
Username & \strut \xmark & \strut \cmark & \strut \cmark & \strut \cmark & \strut \cmark & \strut \cmark & \strut \xmark \\ 
Email & \strut \cmark & \strut \xmark & \strut \xmark & \strut \xmark & \strut \xmark & \strut \xmark & \strut \cmark \\ 
Duration & \strut \xmark & \strut \xmark & \strut \xmark & \strut \xmark & \strut \xmark & \strut \cmark & \strut \xmark \\ 
\bottomrule
 \end{tabu}
	\end{threeparttable}
\end{sidewaystable}

\chapter{Tool descriptions}
\label{ch:descriptions}

This chapter provides short descriptions for all examined tools.
These descriptions focus on aspects that were hard to cover in the presented feature tables, such as the required technical expertise, the assistance for inexperienced form authors or the straightforwardness of the tool's usage workflows.
Additionally, the chapter features a brief summary of another software tool: a suite from the COBWEB-project\footnote{\url{https://cobwebproject.eu}} that was originally supposed to be included in the comparison but was deemed unfit for thorough testing during the process of this student-project. 

\section{EpiCollect5}
EpiCollect5\footnote{\url{https://five.epicollect.net}} is being developed at the Imperial College London and is the successor of Epicollect and Epicollect+ (Plus)\footnote{\url{http://www.epicollect.net}}.
The project gets financial support from the Wellcome Trust Foundation\footnote{\url{https://wellcome.ac.uk}} and has an active community\footnote{\url{https://plus.google.com/u/0/communities/113931398671644426378}} that welcomes questions and allows contact to other projects that use Epicollect as their data collection platform.
Currently, 27.09.18, EpiCollect5 does not allow to register user accounts.
Instead, user authentication is provided by Google, so a Google account is mandatory in order to design and publish a survey on EpiCollect5 and to participate in private surveys.

The tool's workflow is project based: it allows to create a project and then create a form for this project using a clearly arranged form builder.
Projects can be set to private, to allow the project manager to invite users and assign predefined roles (Manager, Curator, Collector) to them.
If, on the other hand, the project is set to be publicly accessible interested people will be able to contribute data to the project and view already collected data, even without having a registered user account. 
Additionally, projects can be hidden so the publicly available project list will not include them as entries - this means only parties that know the project's generated URLs will be able to access them \cite{ec5UserGuideProjectDetails}.

\begin{figure}
    \centering
    \includegraphics[width=1.0\textwidth]{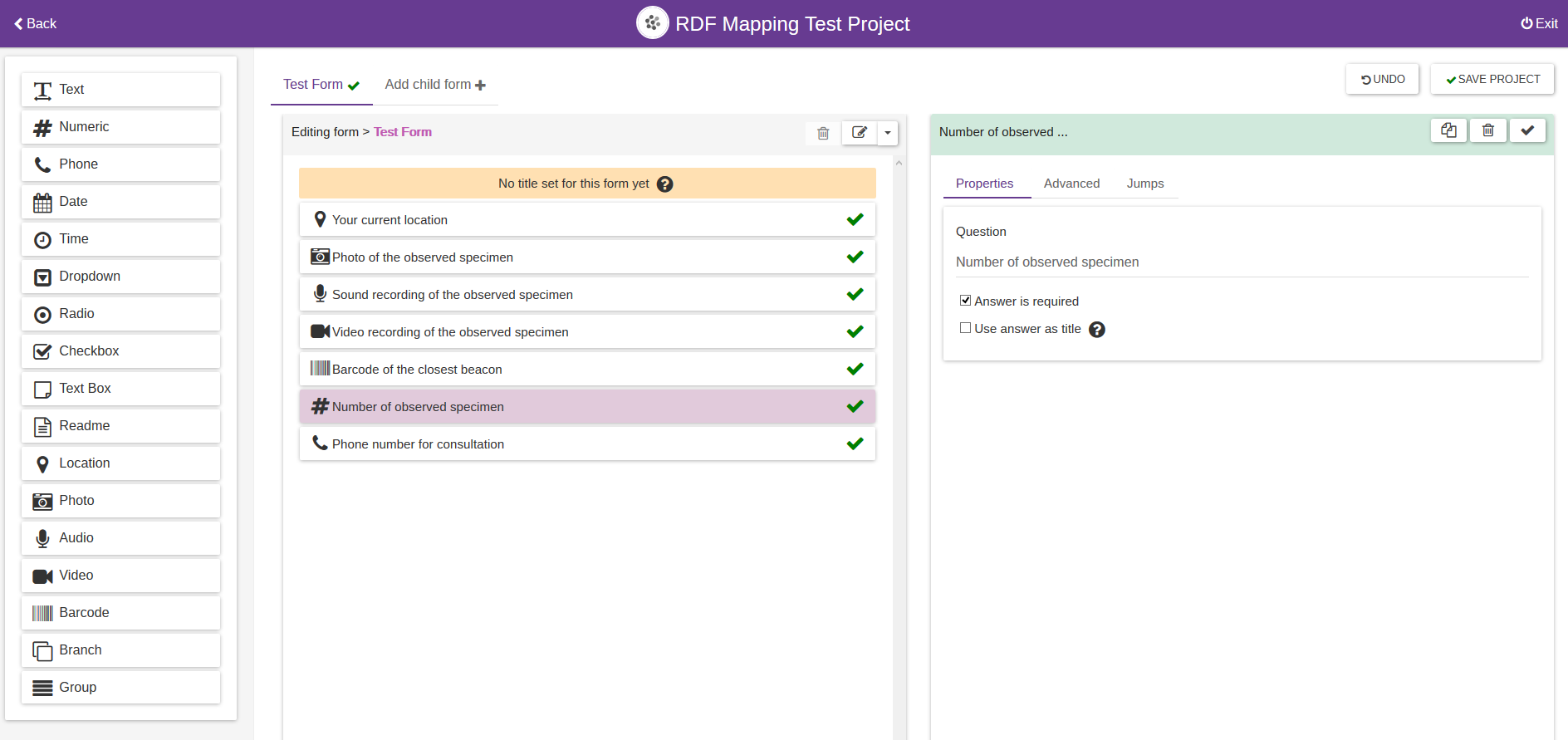}
        \caption{EpiCollect5 - Form builder interface}
        \label{img:formbuilder:epicollect}
\end{figure}

\begin{figure}
    \centering
    \includegraphics[width=0.35\textwidth]{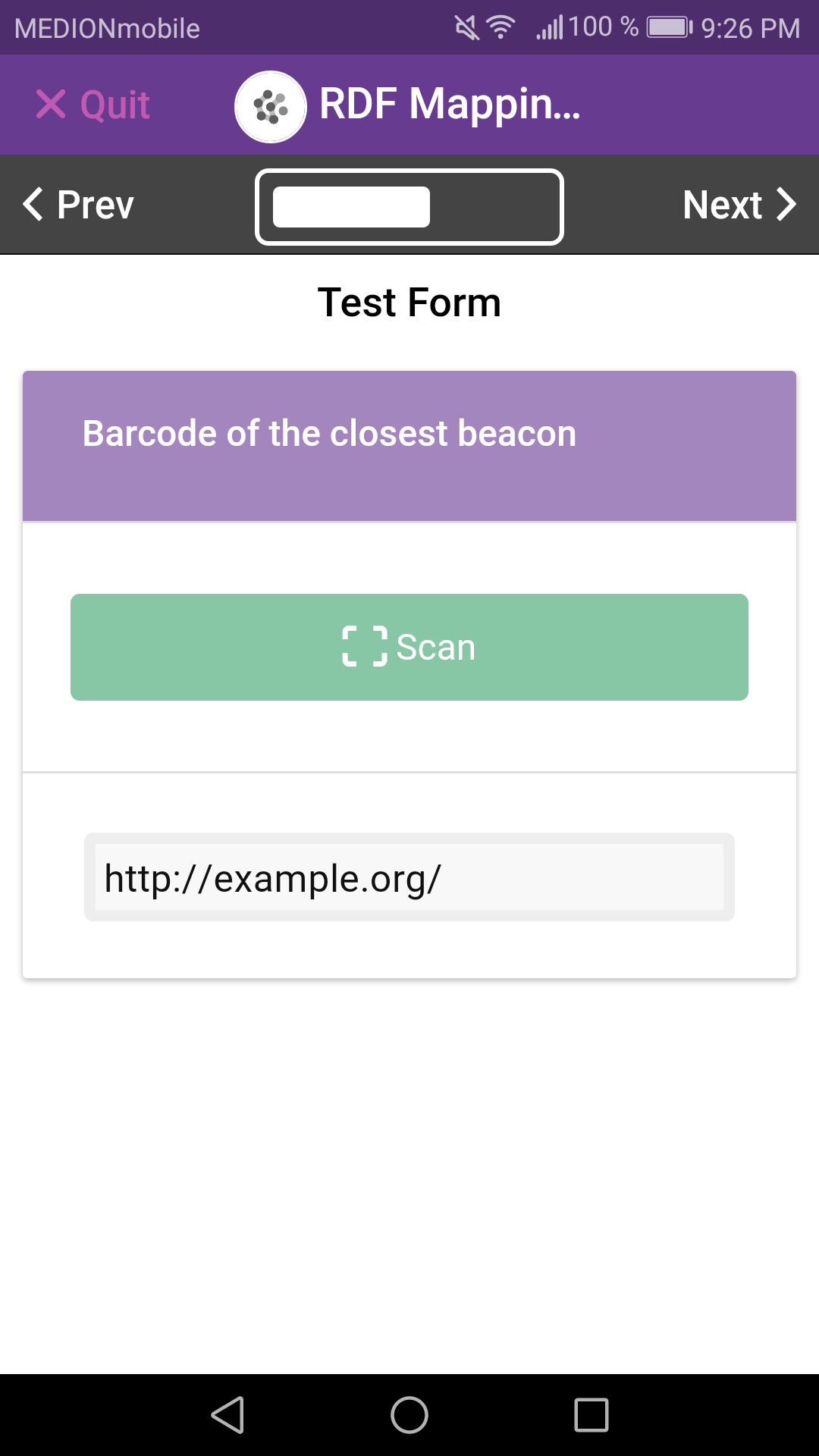}
        \caption{EpiCollect5 - Mobile app}
        \label{img:app:epicollect}
\end{figure}

EpiCollect5 operates entirely web-based, apart from the mobile data collection.
At no stage in a typical project's workflow technical expertise is required to use this tool.
The website is self-explanatory and guides the form author through the steps that are necessary to set up a project, build the form and then collect data using a mobile app.
Figure \ref{img:formbuilder:epicollect} shows the form builder of EpiCollect5.
The app itself is equally easy to use and well designed, Figure \ref{img:app:epicollect} shows one step in a data collection process.
In case any step in the sites usage is unclear, the provided userguide\footnote{\url{https://epicollect5.gitbooks.io/epicollect5-user-guide/content/}} is very detailed and well organized. 

\section{Open Data Kit 1}
\begin{figure}
    \centering
    \includegraphics[width=1.0\textwidth]{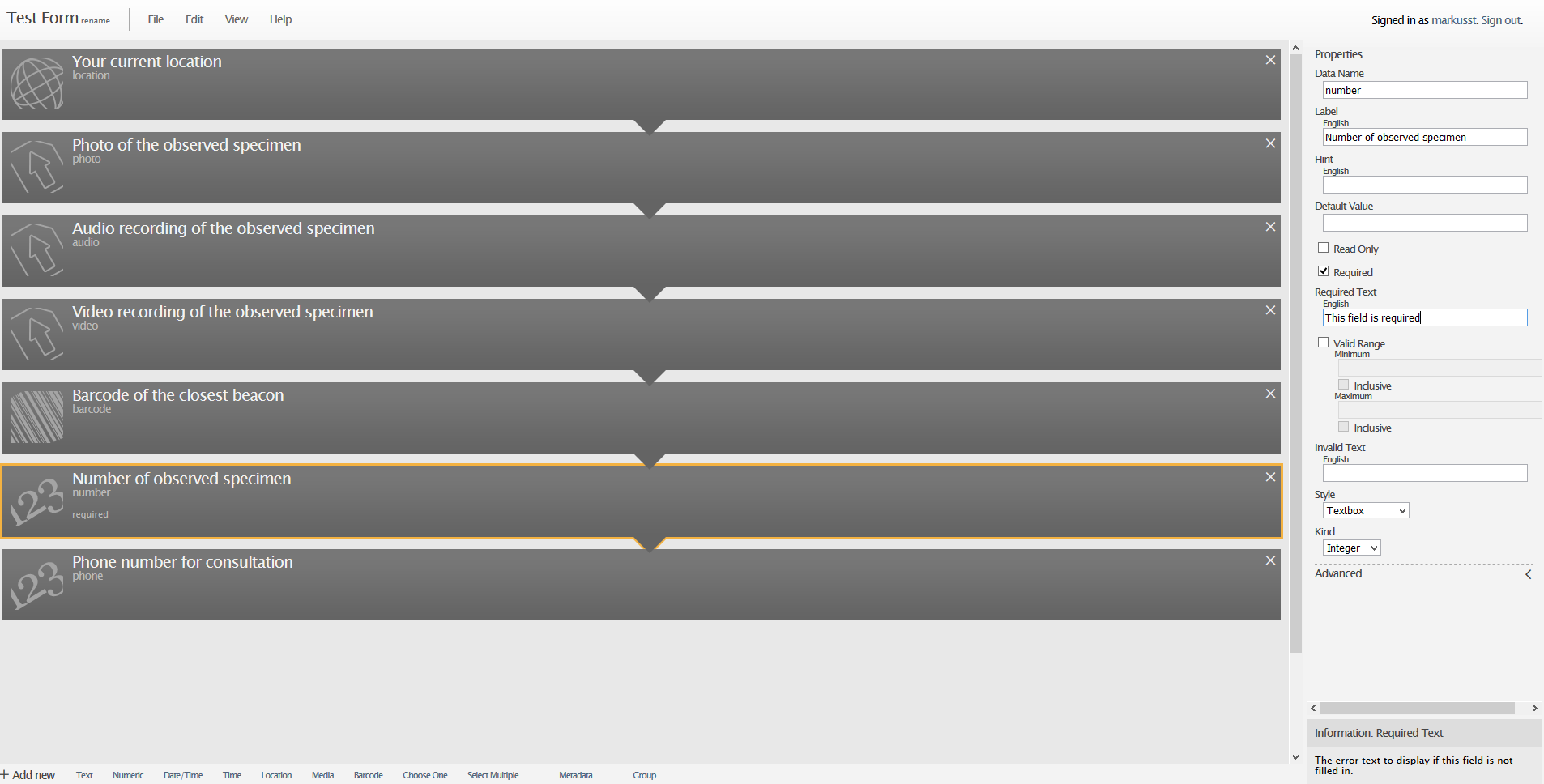}
        \caption{ODK1 - Form builder interface}
        \label{img:formbuilder:odk1}
\end{figure}

Open Data Kit 1 (ODK1)\footnote{\url{https://opendatakit.org/software/}} is one of the two open source tool suits that are being developed by Nafundi\footnote{\url{https://nafundi.com}} and the ODK community. 
Its three main components are ODK Build which is used to create survey forms (Figure \ref{img:formbuilder:odk1}), ODK Aggregate which is used to distribute these surveys and store, export and publish the collected data, and ODK Collect, the Android app that is used for mobile offline data collection.
All of these components can be found on ODKs Github page\footnote{\url{https://github.com/opendatakit}}, along with the communications- and form-description-standards that they are using.

ODK has a very active forum\footnote{\url{https://forum.opendatakit.org}} with enthusiastic members that are eager to help and discuss with ODKs users, developers and scientists.

\begin{figure}
    \centering
    \includegraphics[width=0.35\textwidth]{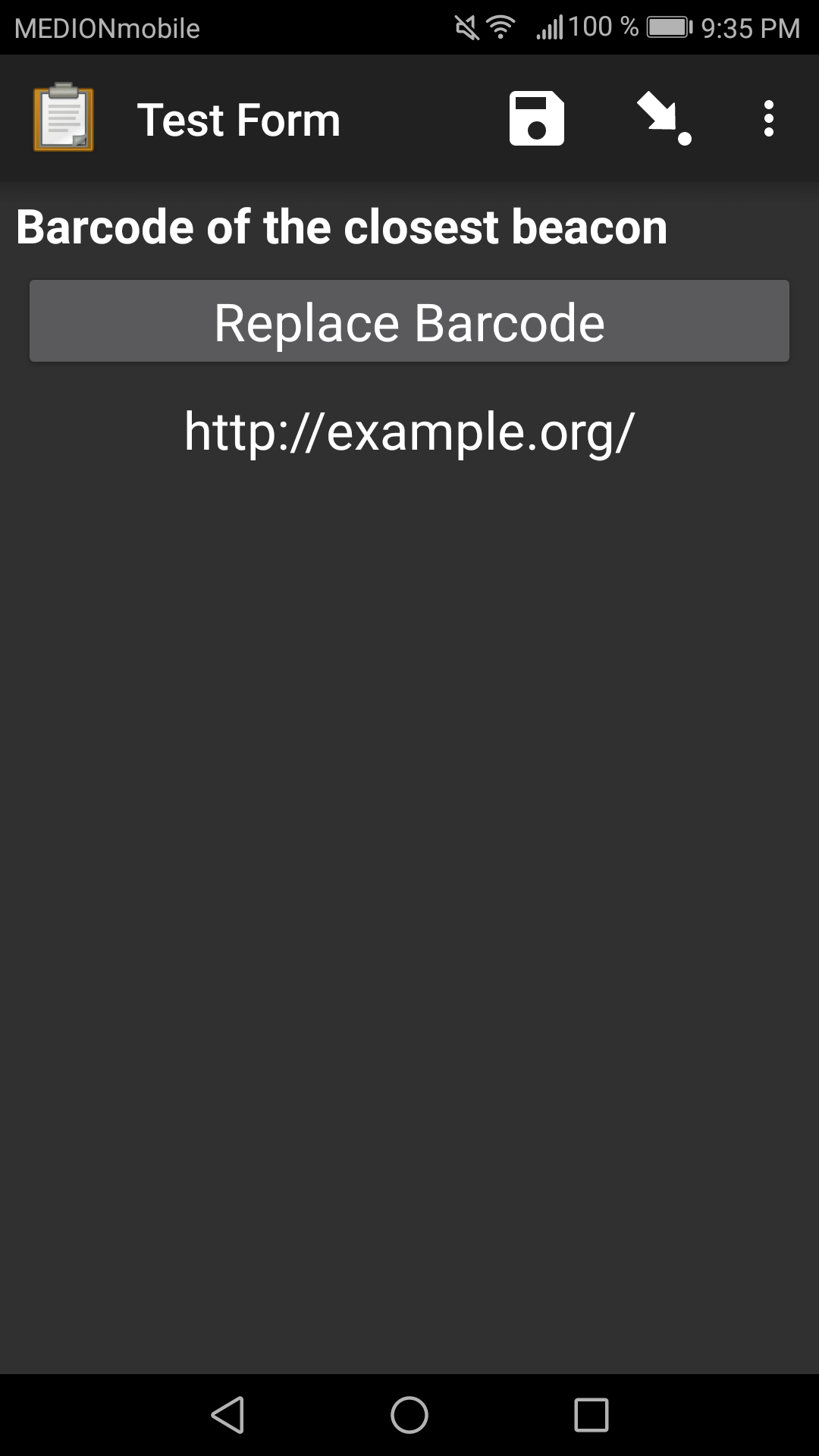}
        \caption{ODK1 - ODK Collect mobile app}
        \label{img:app:odk1}
\end{figure}

In contrast to EpiCollect5, ODK requires a bit of technical knowledge to get started:
While ODK Build is provided as a website\footnote{\url{https://build.opendatakit.org}} and ODK Collect can be found in the Google Play Store\footnote{\url{https://play.google.com/store/apps/details?id=org.odk.collect.android}}, there is only a public sandbox-version for ODK Aggregate\footnote{\url{http://sandbox.aggregate.opendatakit.org}}.
This server is perfect for testing surveys and getting to know ODK, but uploaded forms and data are being deleted regularly.
This means that in order to properly use ODK in a project environment, the project has to set up ODK Aggregate on a cloud platform (Google, AWS, ...) or host the application on its own server or virtual machine.
The ODK documentation provides detailed step-by-step tutorials for all of these options so the required technical knowledge for the setup is minimized \cite{odkAggregateDeployment}. 
Once the server is set up, its URL has to be entered into the ODK Collect app in order to pull surveys and push data to it.
Forms can then be built with ODK Build, pushed to the Aggregate server and then deployed to the mobile app.
All of these steps are described in the ODK Getting Started guide \cite{odkGettingStarted}.
A single step in the data collection process using the mobile app can be seen in Figure \ref{img:app:odk1}

Apart from this technical setup, the usage of the tools is straight forward and intuitive.
Particularly noteworthy is that ODK Build offers explanations for all options that can be used for the various form elements.
This makes it easy even for inexperienced form-developers to create sophisticated surveys and ensure data quality using constraints and similar options.

\section{Open Data Kit 2}
\begin{figure}
    \centering
    \includegraphics[width=0.35\textwidth]{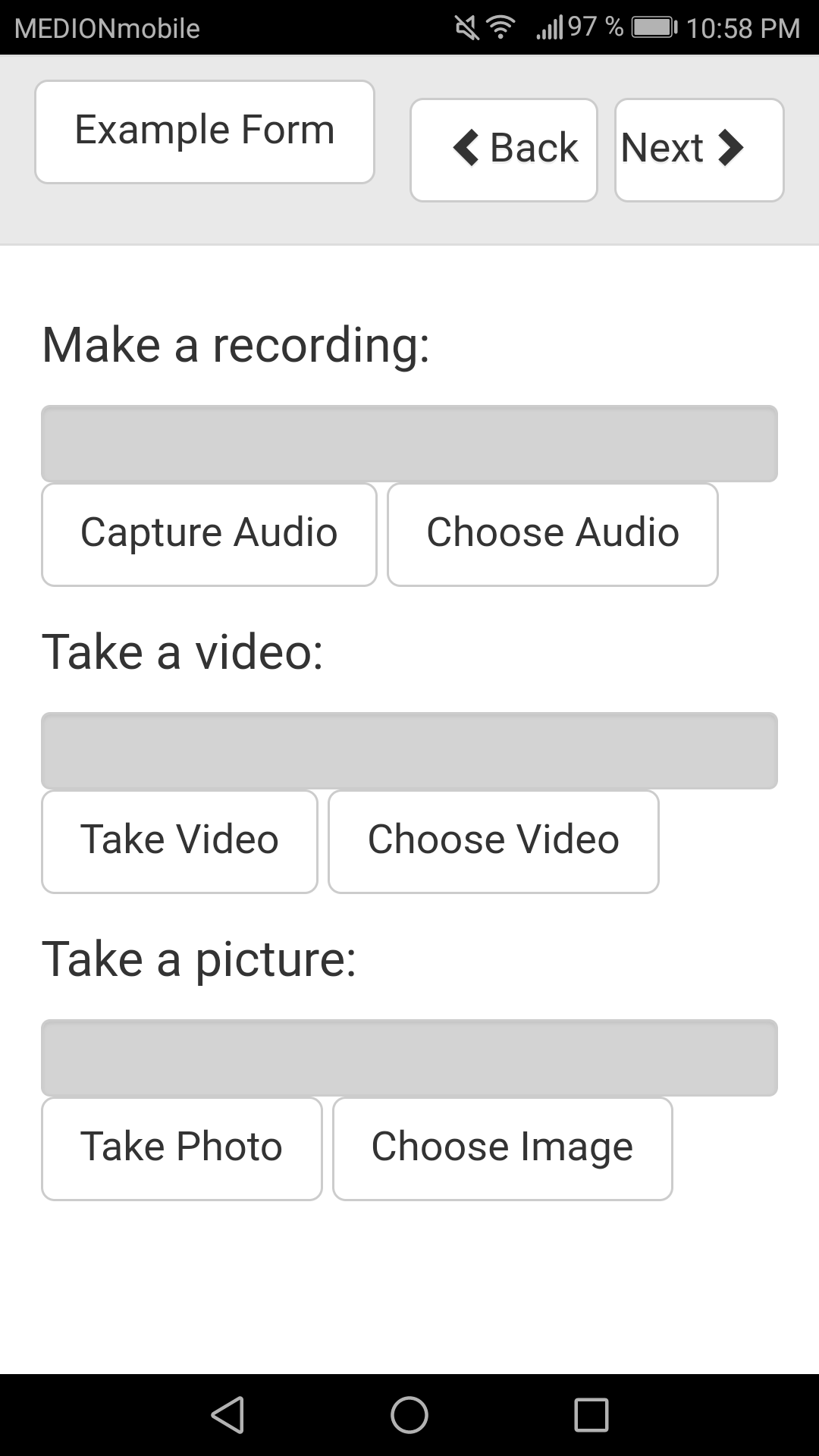}
        \caption{ODK2 - ODK Survey mobile app}
        \label{img:app:odk2}
\end{figure}

Open Data Kit 2 (ODK2)\footnote{\url{https://opendatakit.org/software/}} is the second open source tool suite developed by Nafundi and the ODK community.
It was developed for more elaborate data collection applications than ODK1 and thus it provides additional functionality for the form author, but comes with the downside of a much more technical setup and workflow.
The main components of the suite are the ODK Application Designer that allows to build a data collection application, and ODK Survey (Figure \ref{img:app:odk2}) and ODK Tables, both used for mobile data collection, the latter also for data visualization and management.
In the background, ODK Services is used to handle storage and flow of the data.
All of the components can be found on the ODK Github page\footnote{\url{https://github.com/opendatakit}}.

The community of ODK2 is managed in the same forum as the ODK1 community, though the number of ODK2 users seems to be a lot smaller, judging by the number of topics tagged with ODK2-related keywords ($\approx$60) compared to those tagged with ODK1-related keywords ($>$700) (checked on 24.09.2018).

ODK2 allows to build much more complex data collection applications but it requires a lot of technical knowledge.
A full ODK2 application consists of a rather complex folder-structure that is usually built using the Application Designer.
ODK2 does not provide a form-builder, the form has to be constructed as an XLSForm, an .xls file with a specific structure.
While ODK Build (part of ODK1) allows to export a form in the XLSForm format, not all of its features are compatible with ODK2 and therefore its usage is not always an option.
The XLSForm is then processed by the XLSX Converter of the Application Designer and multiple files are generated and added to the complex folder-structure.
Deploying this application then involves multiple technical steps:
First, an ODK Cloud Endpoint is required.
This can be an extended ODK Aggregate server (see ODK1) or an ODK Sync Endpoint \cite{odk2cloudEndpoint}.
The data collection application then has to be moved to a mobile device and pushed to the server.
Afterwards, other mobile devices should be able to download the application from the server \cite{odk2deployingAnApplication}.

This whole workflow is a lot more technical than in ODK1 and similar tools and even the creation of the form is not as straight forward (due to the lack of a visual form builder).
On the other hand, it allows experienced developers, to build more complex applications by creating custom prompts using HTML, CSS and JavaScript or by utilizing advanced features like queries\footnote{\url{https://docs.opendatakit.org/odk2/xlsx-converter-using/\#using-queries}} or linked tables\footnote{\url{https://docs.opendatakit.org/odk2/xlsx-converter-using/\#linked-tables}} to get data from external resources or modify data in different tables.
The official ODK help page suggests to use ODK2 only if ODK1 does not provide the features required for a certain use case \cite{odkHelp}.

\section{KoBo Toolbox}
KoBo Toolbox\footnote{\url{https://www.kobotoolbox.org}} is an open source data collection tool, quite similar to EpiCollect.
It is mainly being developed by members of the Harvard Humanitarian Initiative (HHI)\footnote{\url{http://hhi.harvard.edu}}.
KoBos mobile app, KoBoCollect, that is used for data collection can be found in the Google play store\footnote{\url{https://play.google.com/store/apps/details?id=org.koboc.collect.android}} and the code for all components can be found on the projects Github page\footnote{\url{https://github.com/kobotoolbox}}, including a Docker\footnote{\url{https://www.docker.com}}-version of the tool that can be used to host the software on any server.
However, since the whole tool is accessible for free on the KoBo website\footnote{\url{https://www.kobotoolbox.org}}, hosting the software yourself is not necessary unless full control over the data storage is required.
If such a self-hosted version is still desired, a detailed step-by-step walkthrough for such a setup is provided in the Github-repository of the Docker-version.

\begin{figure}
    \centering
    \includegraphics[width=1.0\textwidth]{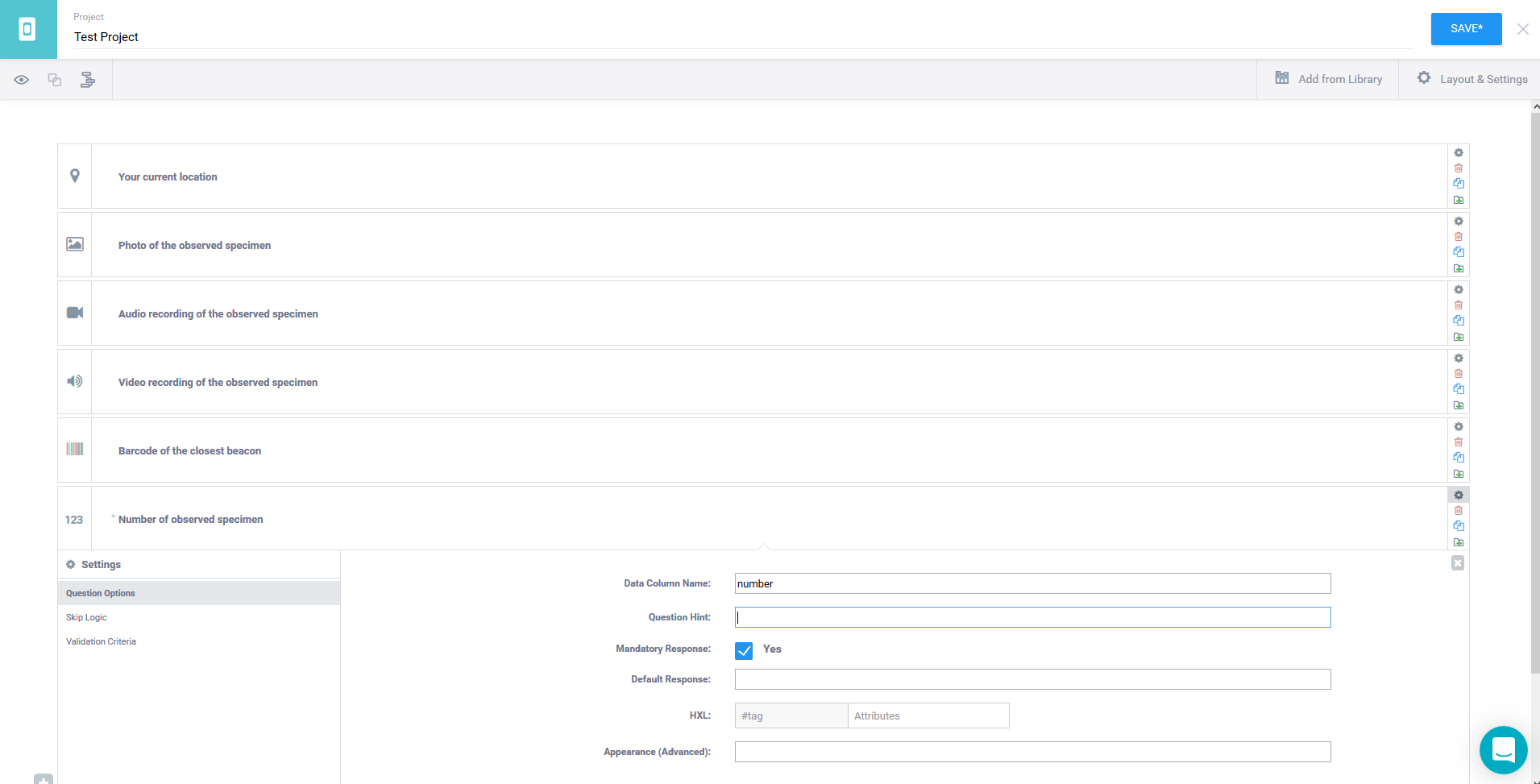}
        \caption{KoBo Toolbox - Form builder interface}
        \label{img:formbuilder:kobo}
\end{figure}

\begin{figure}
    \centering
    \includegraphics[width=0.35\textwidth]{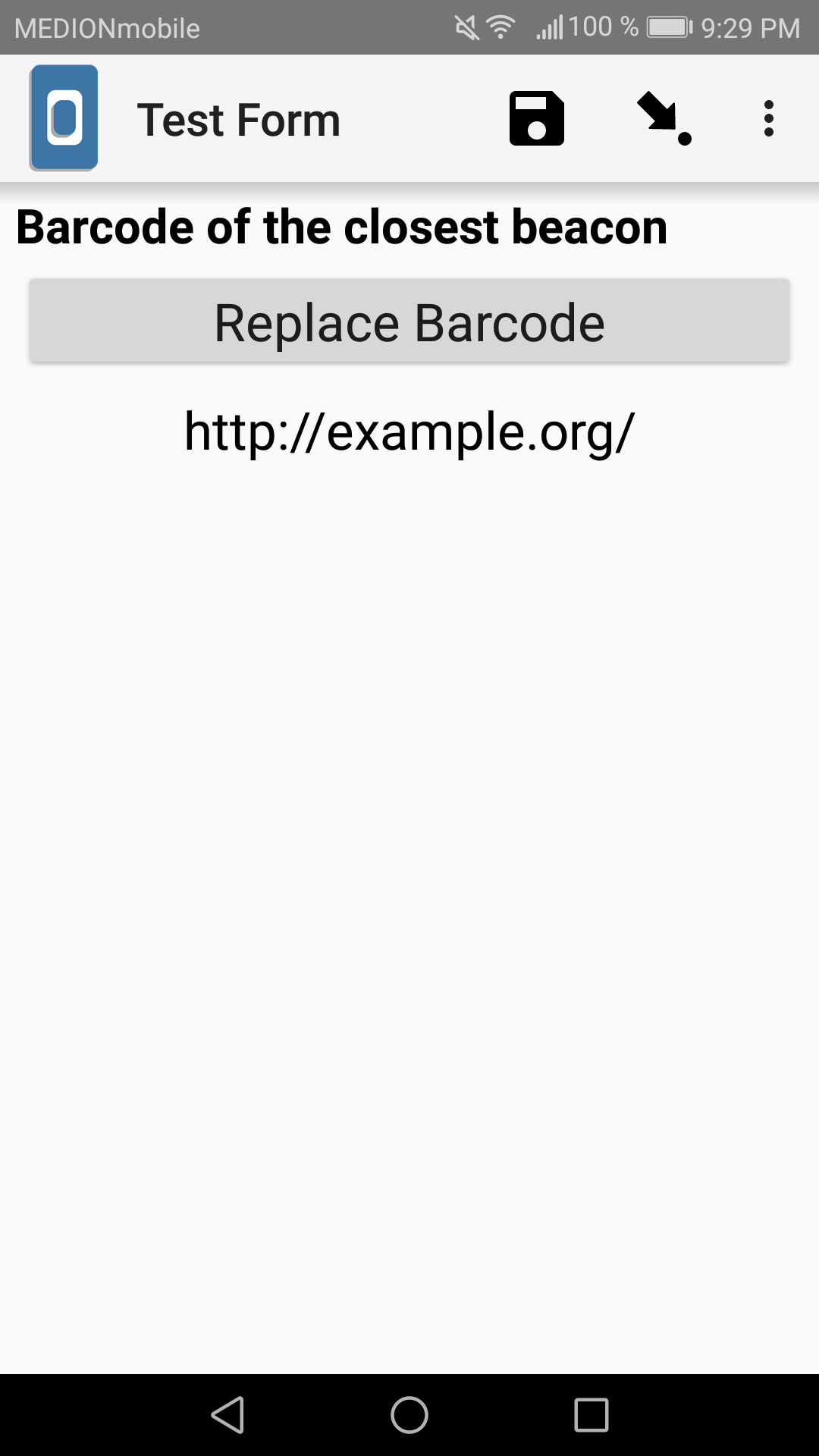}
        \caption{KoBo Toolbox - KoBoCollect mobile app}
        \label{img:app:kobo}
\end{figure}

KoBo works, similar to EpiCollect, in a project-based fashion.
To create a survey, a project is set up and then a form can be built, using the well-designed and intuitive form-builder presented in Figure \ref{img:formbuilder:kobo}.
The project can then be shared publicly or via a generated link or users can be added and provided with read or write access to submissions or the form itself.
All projects that a user has access to are then accessible via the KoBoCollect mobile app.
This app, shown in Figure \ref{img:app:kobo}, is very similar to ODK Collect because it was built on the same codebase.
Alternatively, data can also be collected directly in the web-browser.

KoBo provides a clearly arranged report functionality that can be used to automatically generate and visualize reports, for example to see which percentage of survey participants selected which answers in a multiple-choice question.
The whole workflow is very straight forward and requires no technical knowledge.

\section{Ohmage}
Ohmage\footnote{\url{http://ohmage.org}} is an Open Source data collection platform that promises additional features like data analysis and visualization.
It was developed and maintained by parts of the University of California, Los Angeles and the Cornell Tech school\footnote{\url{https://tech.cornell.edu}}\cite{ohmageAboutUs}.
The code was published on Github\footnote{\url{https://github.com/ohmage/server}} but it seems like the project is not under active development anymore: There has only been a single commit since 2016 which was a rather insignificant change of a single line (checked on 26.09.2018).

Ohmage allows to either host a version of its software or to use a version that is provided by Mobilize\footnote{\url{https://www.mobilizingcs.org}}, an organization that is focused on creating a data science curriculum for students.
A guide for self-hosting is provided in the Github-repository, it only requires the setup of a MySQL database and a Tomcat server \cite{ohmageServerGithub}.
To test the capabilities of the tool in the context of this project, however, the version provided by Mobilize was used.

\begin{figure}
    \centering
    \includegraphics[width=1.0\textwidth]{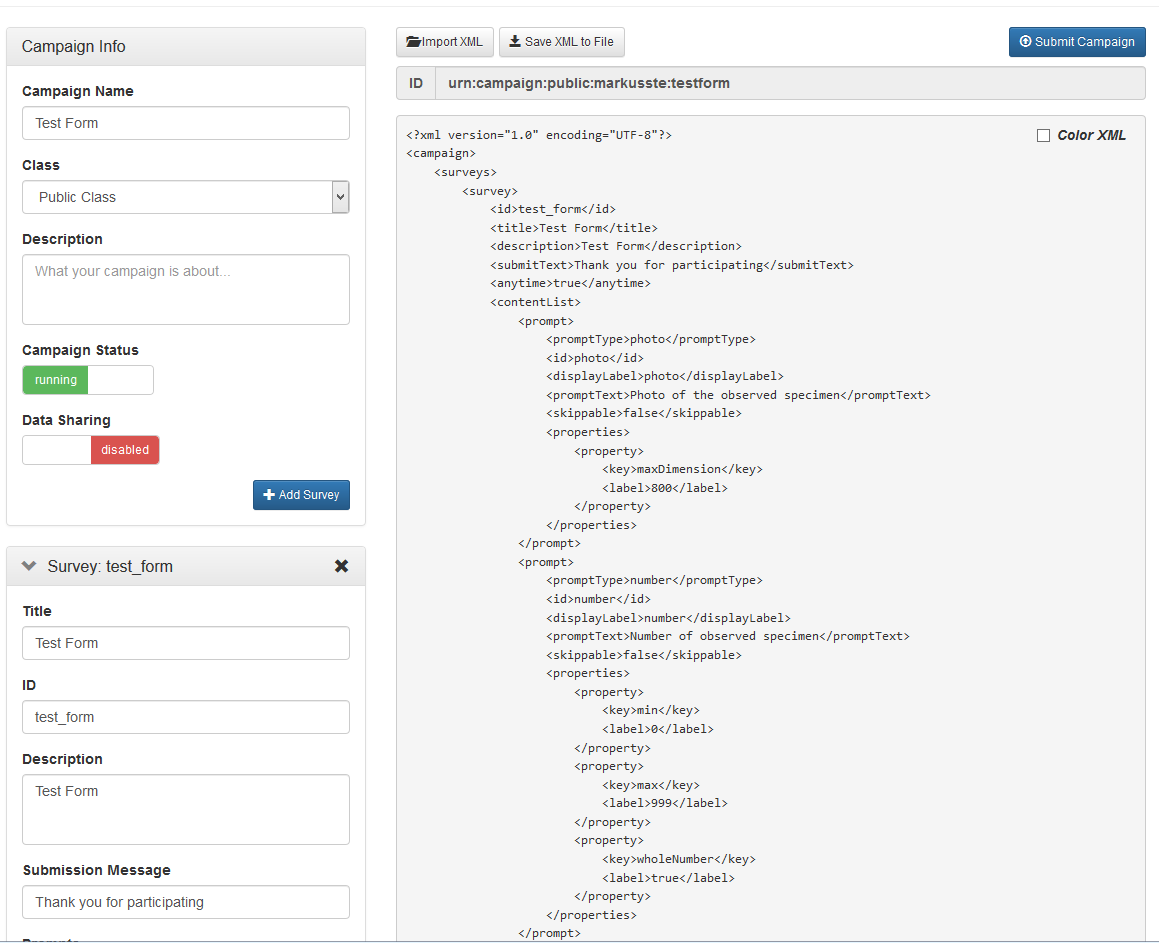}
        \caption{Ohmage - Form builder interface}
        \label{img:formbuilder:ohmage}
\end{figure}

Using the tool feels less intuitive than using the other ones that were examined.
The form author is required to establish a campaign (similar to projects in the other tools) and can then build one or more surveys for the campaign.
The campaign-leader can decide whether or not the data that is collected should be publicly accessible.
The survey itself is built using a rather rudimentary form-builder (Figure \ref{img:formbuilder:ohmage}) that shows the generated XML instead of a visual representation of the survey.
The form-builder often fails to properly explain the options that can be entered for form-elements, for example it is not stated that the numeric "minimum" and "maximum" values refer to the value itself when using a numeric input, but refer to the length of the response when using text input.

\begin{figure}
    \centering
    \includegraphics[width=0.35\textwidth]{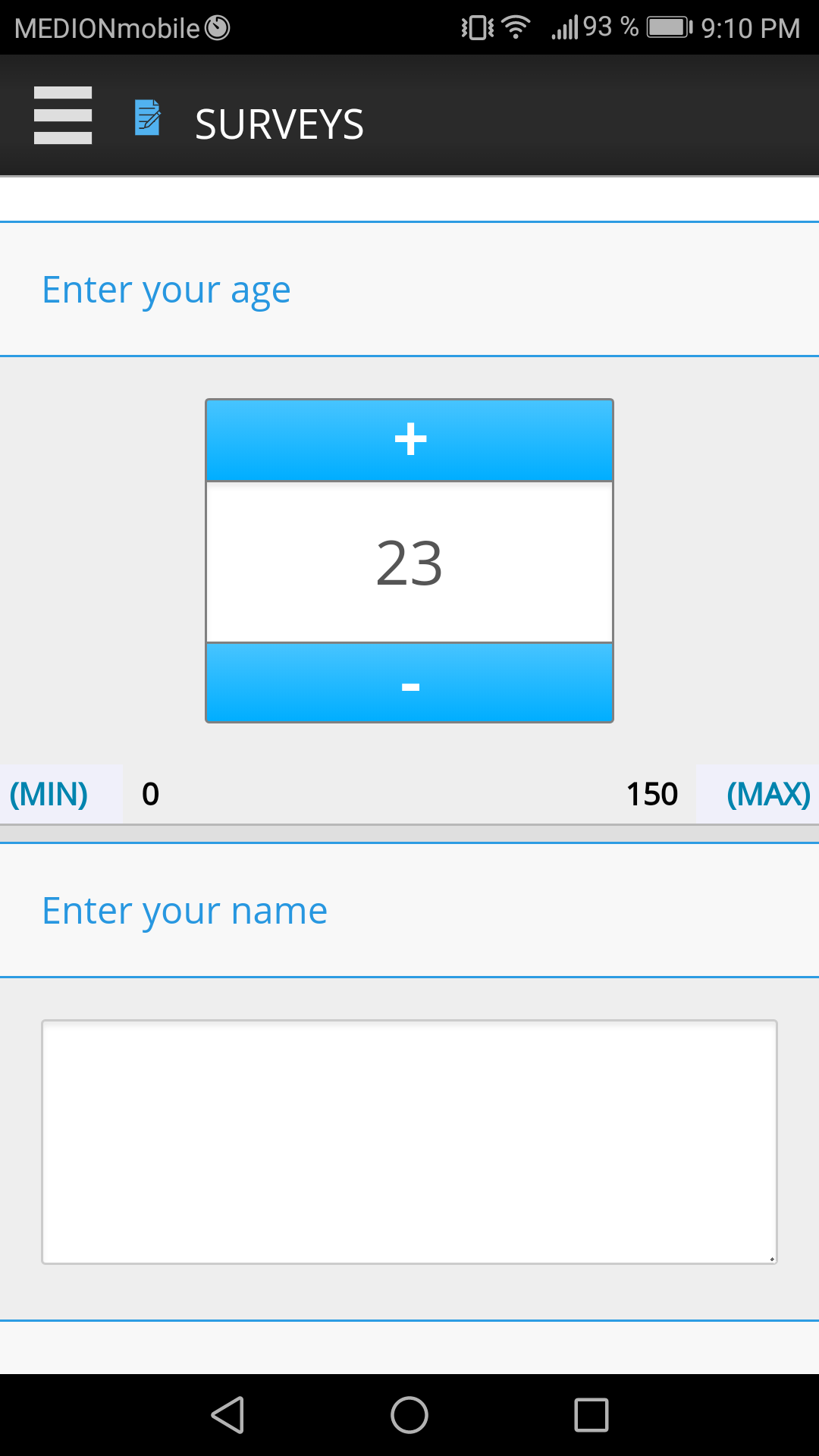}
        \caption{Ohmage - UCLA MobilizingCS App}
        \label{img:app:ohmage}
\end{figure}

After publishing the campaign, the survey can be filled either in the web browser or via mobile device using the UCLA MobilizingCS App\footnote{\url{https://play.google.com/store/apps/details?id=org.ohmage.mobilizingcs&hl=en}} visible in Figure \ref{img:app:ohmage}.
Uploaded responses can afterwards be visualized and analyzed using different tools: an interactive dashboard to display charts and maps for data and metadata, a plot app to generate plots of data or a monitoring tool to visualize metadata regarding responses and users.

Overall, while the built-in visualization is a nice feature, Ohmage doesn't offer much when compared with other open source tools like EpiCollect and Open Data Kit.
The workflow on the website is confusing, only basic form-elements are available to create a survey and the fact that the tool is not under active development anymore diminishes the hope for improvement.
So if the built-in visualization (which can easily be replaced by more specialized tools) is not an absolute must have, I would not recommend Ohmage as a data collection platform.
\section{SurveyCTO}
SurveyCTO\footnote{\url{https://www.surveycto.com}} is a paid subscription-based data collection platform which also offers a free "Community Subscription".
This option comes with some rescrictions regarding the number of teams, forms and submissions per month and requires monthly feedback, often in the form of a survey, to keep the subscription active.
The platform offers a very clean and professional website that requires no technical expertise and is very user-friendly.
It was developed by Dobility, a company that has declared its goal to "increase the quality of data, research, and analysis by providing affordable technology that anyone can use"\cite{dobility}.
The tool has since been utilized by different major organizations like the Johns Hopkins Bloomberg School of Public Health (JHSPH)\footnote{\url{https://www.jhsph.edu}} and the Clinton Health Access Initiative\footnote{\url{https://clintonhealthaccess.org}}\cite{dobility}.

\begin{figure}
    \centering
    \includegraphics[width=1.0\textwidth]{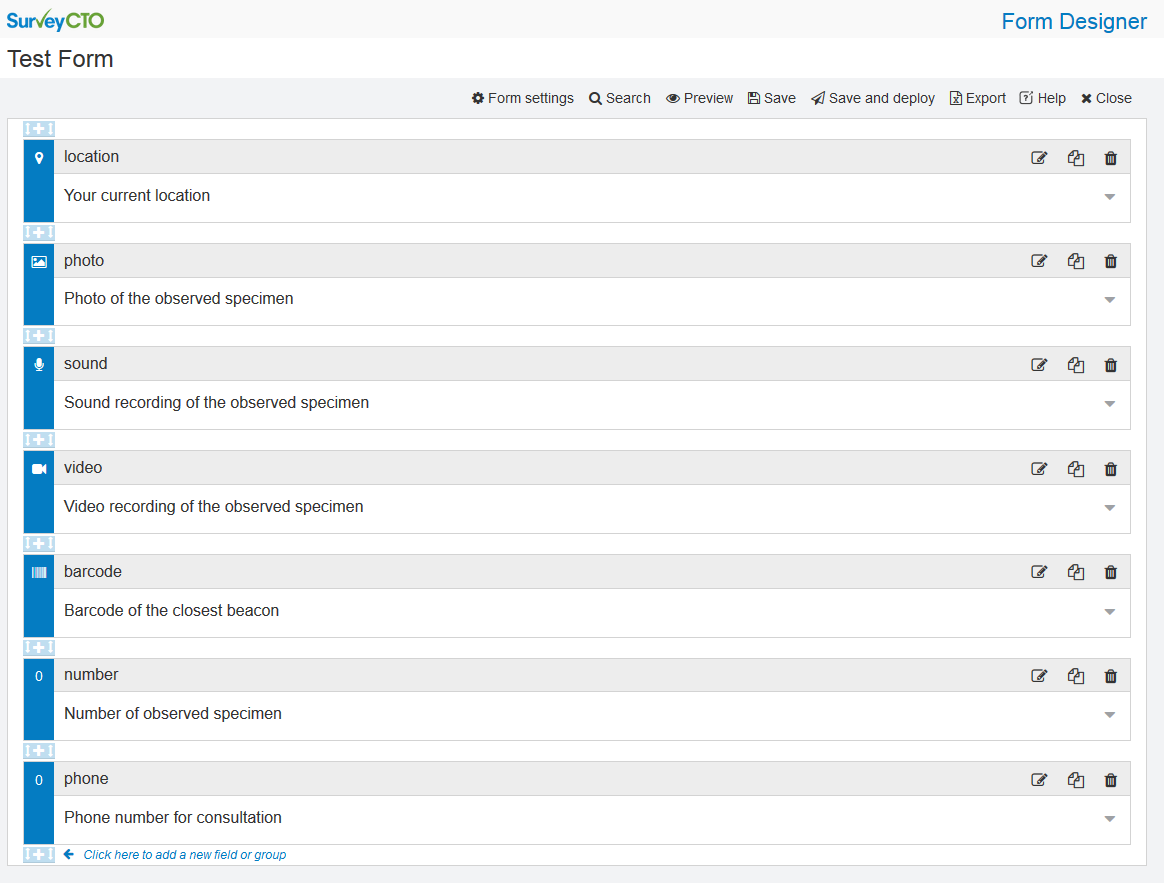}
        \caption{SurveyCTO - Form builder interface}
        \label{img:formbuilder:surveycto}
\end{figure}

The first step in the usual workflow of the tool is to create a virtual server for the form author.
This is done by simply entering some basic information, the creation of the server itself is completely transparent.
The created server is then used to manage forms, submitted data and users.
SurveyCTO puts great emphasis on a user-friendly workflow:
Forms are built using a very extensive form-builder, visible in Figure \ref{img:formbuilder:surveycto}, and the form author can choose from a number of form-templates that showcase lots of form-elements and their configuration.
Extensive help-texts are provided that explain all elements of the workflow to inexperienced authors.
Additionally, wizards are provided that help to generate constraints or skip-logic - this is especially helpful for authors with little knowledge of complex logic statements.

\begin{figure}
    \centering
    \includegraphics[width=0.35\textwidth]{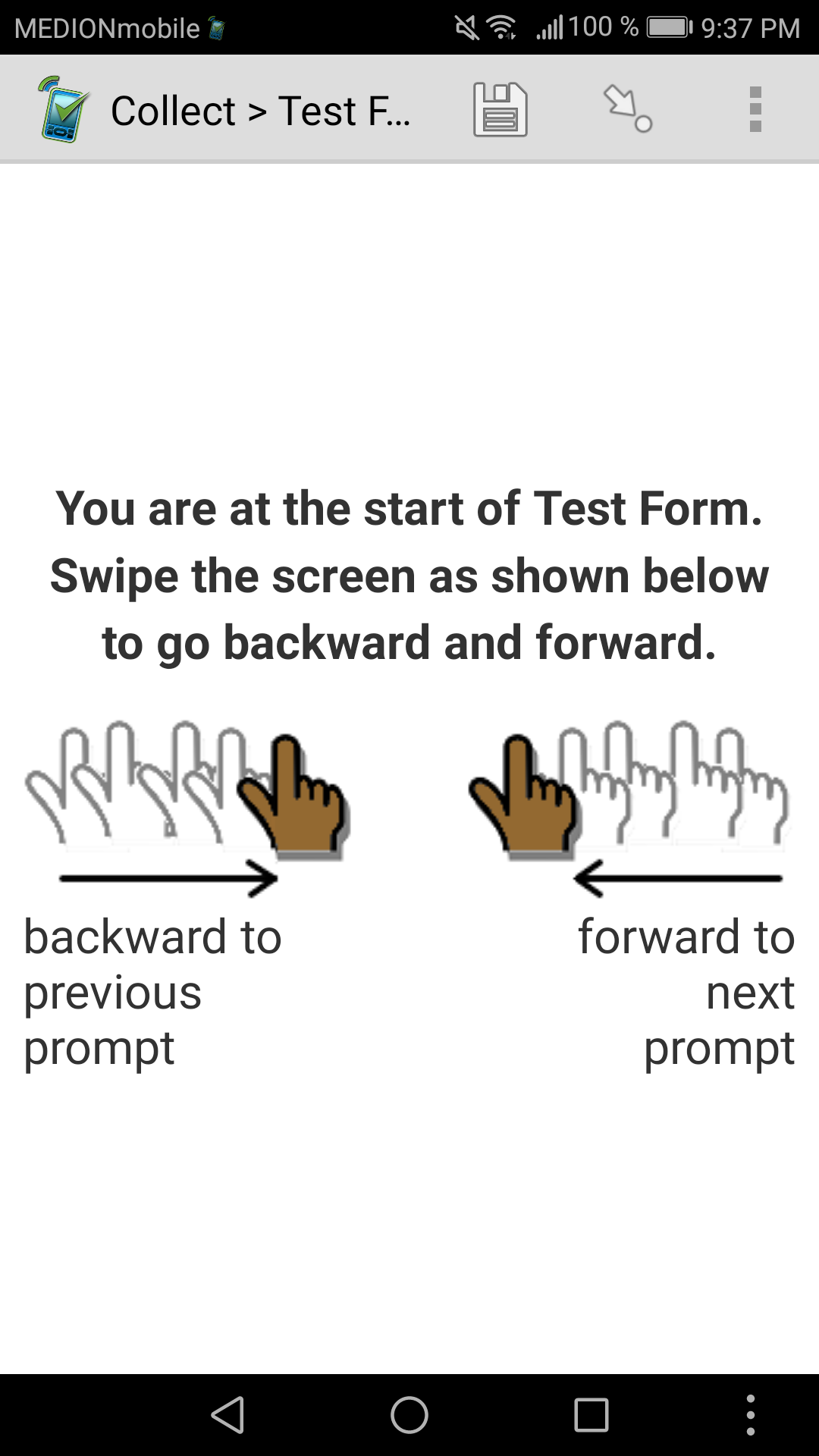}
        \caption{SurveyCTO - SurveyCTO Collect}
        \label{img:app:surveycto}
\end{figure}

Once the form is completed, data can either be collected using the simple but intuitive mobile app SurveyCTO Collect\footnote{\url{https://play.google.com/store/apps/details?id=com.surveycto.collect.android}} shown in Figure \ref{img:app:surveycto} or web data collection can be enabled.
Users that have access to the survey are managed in a separate part of the website where users can be added and predefined roles with specific rights can be assigned to them.

Collected data can be explored on the website using the SurveyCTO Data Explorer that allows simple visualization, for example in the form of bar charts or pie charts.
An additional feature that can drastically improve data quality are automated quality checks, in addition to the validation on the collection device.
Such checks can be defined in a wizard that allows to detect submitted values that are outliers, values that occur too frequently, ...
The quality check can then be configured to run regularly and to report the results to an email address.

Data that is stored on the SurveyCTO server can either be downloaded directly as a CSV file or the SurveyCTO Sync software can be installed to manage data export.
This tool allows to download the data from the server and store it locally.
It provides support for CSV, XLSX, KML for geographical information and an export to Stata\footnote{\url{https://www.stata.com}}-compatible files.
It also allows to directly publish data on cloud-based services like Google Sheets or Google Fusion Tables.

Overall, SurveyCTO is a very well designed and professional platform with an intuitive workflow and extensive descriptions for inexperienced form authors.
Small scale projects that are able to deal with the restrictions of the community subscription should definitely have a look at this tool.
Projects that include highly sensitive data might be reluctant to use SurveyCTO because it does not include a license to host the software on a private server.
However, for such cases data encryption can be used to make sure no one will be able to access the data.
In the rare case that this feature does not offer the required security, self-hosting can be organized on a case-by-case basis according to SurveyCTOs support, provided that incurring costs will be settled by the project.
\section{Magpi}
Magpi\footnote{\url{https://home.magpi.com}} is a data collection platform that was founded in 2003 \cite{magpiAbout} and has since been supported by major organizations like OXFAM\footnote{\url{https://www.oxfam.org}}, the World Health Organization (WHO)\footnote{\url{http://www.who.int}} and unicef\footnote{\url{https://www.unicef.org}}.
Similar to SurveyCTO, Magpi has a paid subscription model but also offers a free subscription with several restrictions, for example: the number of forms, questions per form and data submissions are restricted, the number of data collectors is capped and the form-elements for photo-, barcode- and signature-questions are not available.

\begin{figure}
    \centering
    \includegraphics[width=1.0\textwidth]{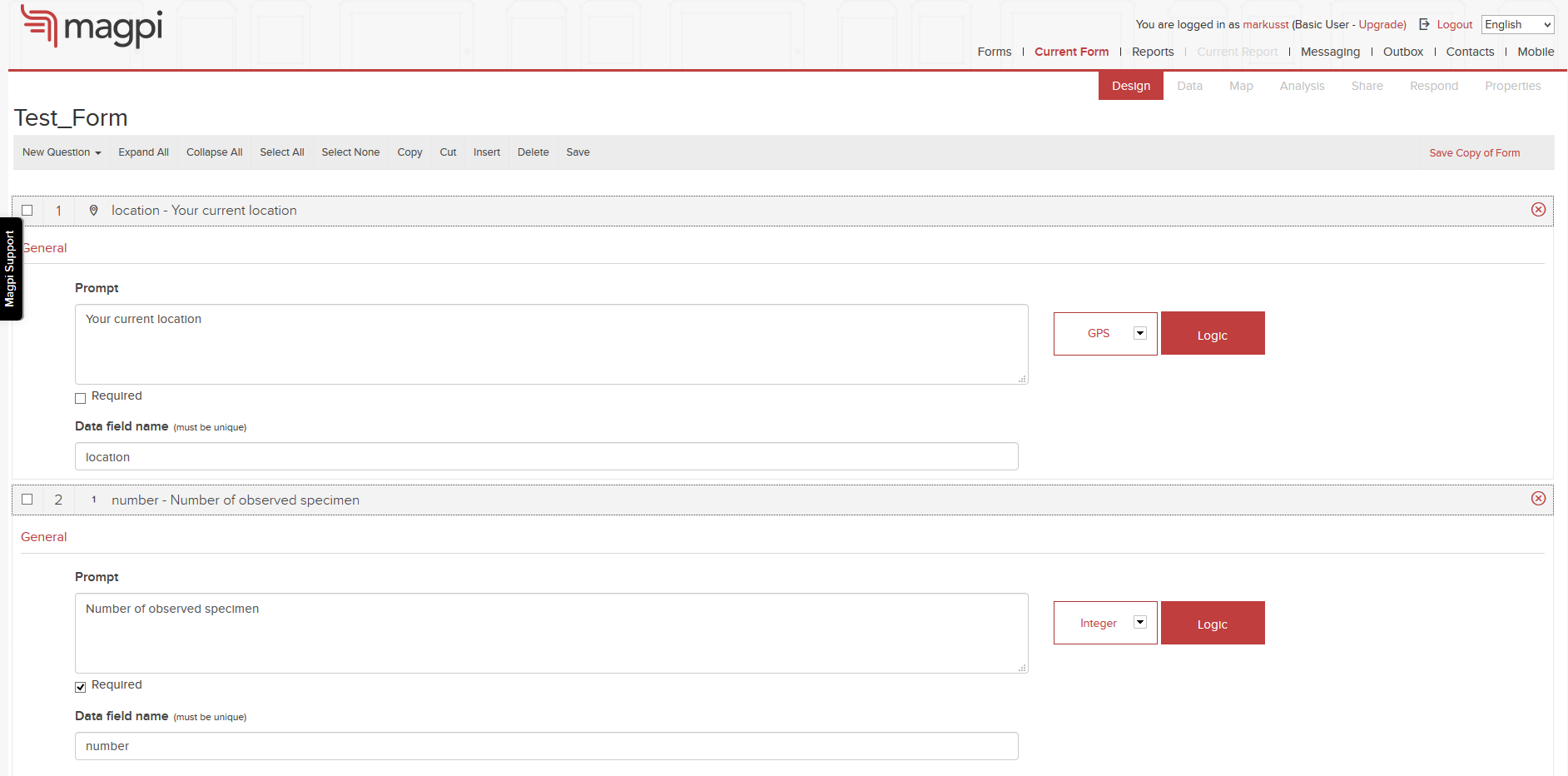}
        \caption{Magpi - Form builder interface}
        \label{img:formbuilder:magpi}
\end{figure}

\begin{figure}
    \centering
    \includegraphics[width=0.35\textwidth]{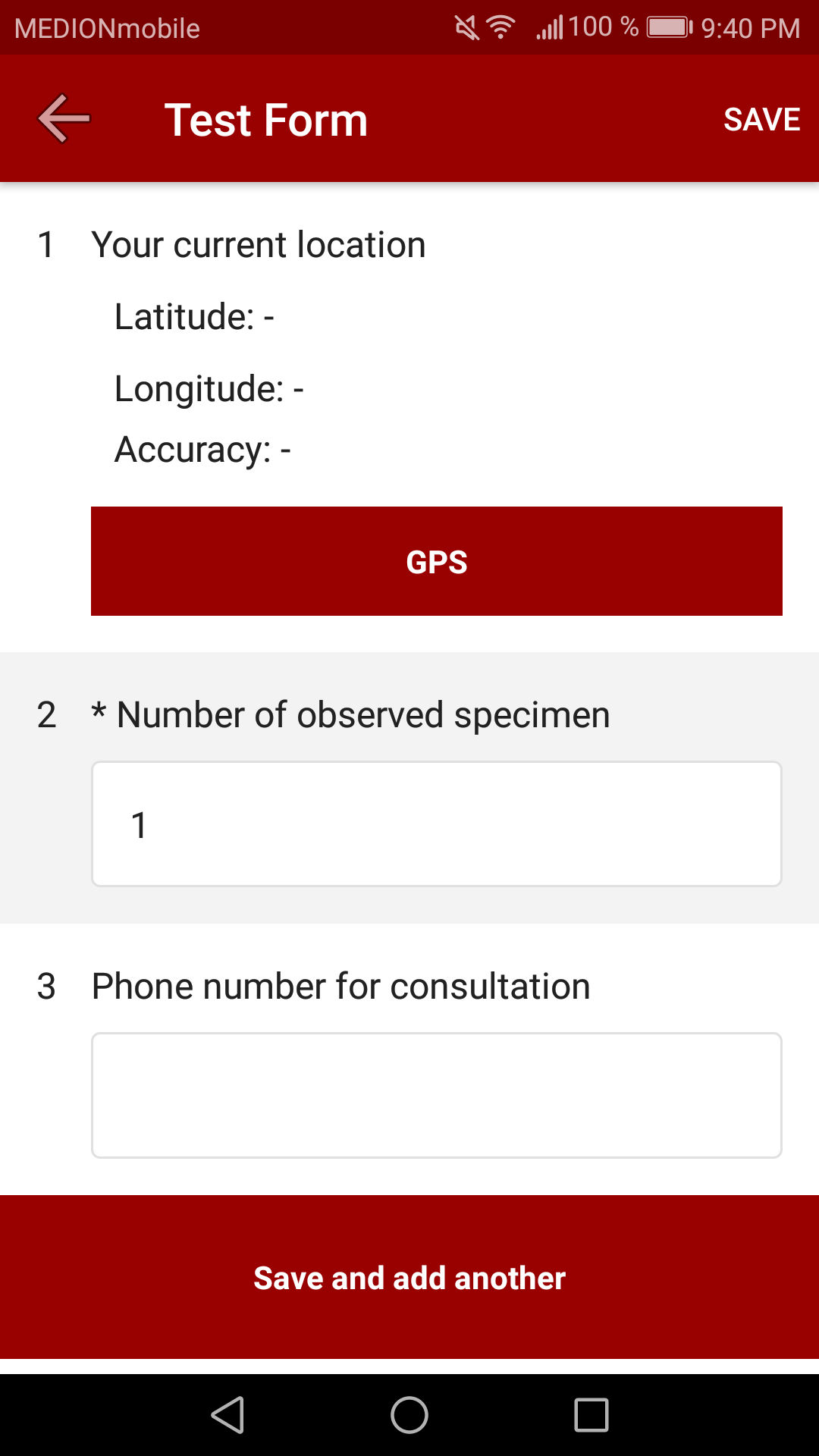}
        \caption{Magpi - Magpi+}
        \label{img:app:magpi}
\end{figure}

Since Magpi is not a project-based tool like, for example, EpiCollect5, the form author can start to design a form right away.
The form-designer offers a small set of public templates that can be copied and adapted to the authors needs.
While this is perfect for new authors, the set of provided templates is a lot smaller than in SurveyCTO.
Figure \ref{img:formbuilder:magpi} shows the interface of Magpi's form builder.
The tool also offers assistance to build complex logic statements for skip logic and calculations.
Mobile data collection is provided by the Magpi+ app\footnote{\url{https://play.google.com/store/apps/details?id=com.magpi.nsapp&hl=en}}, a well-designed app with an intuitive user interface, shown in Figure \ref{img:app:magpi}.
While Magpi does offer a report and visualization tool to its paying users, this feature is not accessible for free.

Overall, Magpi is a little less intuitive and user-friendly than SurveyCTO and offers less features when comparing the free versions of both tools.
However, it is a well-designed tool that requires no technical knowledge and provides a very well-designed mobile application.

\section{COBWEB}
The Citizen Observatory Web project (COBWEB)\footnote{\url{https://cobwebproject.eu}} was an EU-project that focused on advancing citizen science projects that involve mobile data collection.
The project was active from 2012 to 2016 \cite{cobwebWebsite} and was led by the University of Edinburgh in cooperation with multiple European countries \cite{cobwebProjectDetails}.
While the program was mainly focused on the domain of biology and biodiversity, the resulting software should have been applicable in other domains as well.

The developed software is split into multiple parts.
Fieldtrip-Open\footnote{\url{https://github.com/edina/fieldtrip-open}} is the main tool that is used for data collection.
The dedicated website\footnote{\url{http://fieldtrip.edina.ac.uk}} of this tool that is referenced in \cite{cobwebWebsite} seems to be no longer accessible.
However, the code can still be found on Github.
To develop surveys that can be used in Fieldtrip-Open, a survey designer\footnote{\url{https://github.com/edina/survey-designer}} was developed.
This tool requires an additional custom software, the Personal Cloud API (PCAPI)\footnote{\url{https://github.com/cobweb-eu/pcapi}}, to run in the background as a middleware that provides access to cloud-storage software like Dropbox.
The last piece of software that belongs to this suite is Viewer\footnote{\url{https://github.com/cobweb-eu/viewer}}, a tool that allows to visualize geographical data on a map.

Due to the missing documentation for all of these tools, the software could not be fully set up in order to properly test its capabilities in the scope of this student project.
Multiple attempts to set PCAPI up on a Windows OS were unsuccessful, first using Python which resulted in old dependencies being rejected, thus aborting the installation, and later using Docker not resulting in error messages but neither in a running PCAPI instance exposed on the given port.
PCAPI and the survey designer could finally be set up on a Linux based computer, but were deemed unfit for proper testing:
The button that is supposed to save the designed survey resulted in nothing but JavaScript errors hinting at (1) access control problems and (2) problems inside the running PCAPI instance.
It is unclear whether these errors occured due to software problems or due to incorrect or missing configuration (which, in turn, could be the result of missing documentation).
The setup of the main software, Fieldtrip-Open was unsuccessful due to the lack of proper documentation as well.
Since the dedicated website for Fieldtrip-Open is not available anymore, the only source of guidance is the fragmentary and sketchy installation guide\footnote{\url{https://github.com/edina/fieldtrip-open/wiki/install}} in the projects Github-wiki.
This guide, however, was not enough to successfully set up the project, it seems to require some internal knowledge about the project.

Some of the intended features of the software can be derived from published COBWEB reports and other documents.
The example configuration on the Github page of the survey designer shows that the following form-elements seem to be available:
\begin{itemize}
\item Text
\item Range
\item Textarea
\item Checkbox
\item Radio
\item Select
\item Image
\item Audio
\item Warning
\item Dtree
\end{itemize}

\cite{cobwebPublicDeliverables} mentions some features that were planned for the data collection software:
The software was intended to be used across different mobile operating systems \cite[Section 4.1]{cobwebPublicDeliverables}.
Data quality was to be ensured by calculating a "statistical compatibility" with knowledge bases and models that have to be provided prior to the data collection \cite[Section 4.2]{cobwebPublicDeliverables}.
\cite[Section 3.4]{cobwebPlatformRelease} mentions even more ways of ensuring data quality: "location based services, cleaning, automatic validation, comparison with authoritative data, model based validation, big/linked data, and semantic harmonisation".
According to \cite[Section 3.5]{cobwebPlatformRelease} a quality assurance designer was developed in order to support the design of such quality assurance models.
This piece of software, even though it can be found on COBWEBs Github page\footnote{\url{https://github.com/cobweb-eu/cobweb-qa}}, is not part of the list of software outputs presented in \cite{cobwebSoftwareOutputs}.
This most likely means that it was never developed to a point where it is ready for a public release.

In \cite[3.8]{cobwebPlatformRelease} it is stated that data access is provided using the SPARQL Protocol and RDF Query Language (SPARQL), among other query languages.
This access is made possible by serializing the collected data into the necessary format using a provided template.

Overall, the COBWEB software for mobile data collection is currently not in a state that allows proper usage.
Nevertheless, the published reports hint at a strong focus on data quality assurance and even support for semantic web technologies like RDF and access via SPARQL, which sounds very promising.
However, since the project ended in 2016 the software is no longer in active development and it remains to be seen if someone will pick it up again in the future.

\chapter*{Conclusion}
\addcontentsline{toc}{chapter}{Conclusion} 
The presented comparison shows that due to the different features that are offered by the different software tools, the choice of a platform depends on the given use case with its unique requirements.
However, it also shows that ODK1 and KoBo Toolbox are the open source tools that offer the biggest collections of features.
SurveyCTO on the other hand offers the most professional and user-friendly environment if the limitations of the free subscription are not a problem.
For most data collection projects, at least one of these three tools should be able to cover the requirements.

Another point that this comparison shows very clearly is that none of the tools currently provide any kind of semantic component that would allow direct interpretation of the exported data.
The only tool that seemed to take a step in this direction was the COBWEB software suite with its RDF export, which was not fit for proper testing.
Since linked and semantically enriched data enable more meaningful interpretation of the data and even automated reasoning, such features could drastically improve both quality and usability of the collected data.
Therefore, such features deserve some attention in the future enhancement of data collection platforms and tools.

%% ----------------------------------------------------------------
% Now begin the Appendices, including them as separate files

\addtocontents{toc}{\vspace{2em}} % Add a gap in the Contents, for aesthetics

\appendix % Cue to tell LaTeX that the following 'chapters' are Appendices

\addtocontents{toc}{\vspace{2em}}  % Add a gap in the Contents, for aesthetics
\backmatter

%% ----------------------------------------------------------------
\label{Bibliography}
\lhead{\emph{Bibliography}}  % Change the left side page header to "Bibliography"
%\bibliographystyle{unsrtnat}  % Use the "unsrtnat" BibTeX style for formatting the Bibliography
  % The references (bibliography) information are stored in the file named "Thesis.bib"
\printbibliography

\end{document}